\documentclass[aps,prb,twocolumn,showpacs,amsmath,amssymb,floatfix,longbibliography,nofootinbib,superscriptaddress,10pt]{revtex4-2}
\usepackage{graphicx,mathtools,bm,array,enumitem,overpic,booktabs,multirow}
\usepackage{natbib}
\usepackage[dvipsnames]{xcolor}
\usepackage{braket}
\usepackage{float}
\usepackage{textcomp}
\usepackage{comment}
\usepackage[bookmarks=true,colorlinks,linkcolor=RoyalBlue,urlcolor=RoyalBlue,citecolor=RoyalBlue]{hyperref}
\usepackage{cleveref}
\usepackage{hhline}
\usepackage[normalem]{ulem}

\newcommand{\PR}{P}

\newcommand{\sx}{{\tilde{\sigma}}^x}
\newcommand{\sy}{{\tilde{\sigma}}^y}
\newcommand{\sz}{{\tilde{\sigma}}^z}
\newcommand{\spp}{{\tilde{\sigma}}^+}
\newcommand{\spm}{{\tilde{\sigma}}^{\pm}}
\newcommand{\smm}{{\tilde{\sigma}}^-}
\newcommand{\sxN}{\sigma^x}
\newcommand{\syN}{\sigma^y}
\newcommand{\szN}{\sigma^z}
\newcommand{\spN}{\sigma^+}
\newcommand{\smN}{\sigma^-}
\newcommand{\Jp}{J^+}

\newcommand{\Jm}{J^-}
\newcommand{\Jx}{J^x}
\newcommand{\Jy}{J^y}
\newcommand{\Jz}{J^z}
\newcommand{\sr}{I}
\newcommand{\da}{{\downarrow}}
\newcommand{\ua}{{\uparrow}}
\newcommand{\id}{\mathbf{1}}
\newcommand{\GSy}{{\ket{{\rm GS}_y}}}
\newcommand{\GSz}{{\ket{{\rm GS}_z}}}
\newcommand{\CSy}{{\ket{{\rm CS}_y}}}
\newcommand{\CSz}{{\ket{{\rm CS}_z}}}
\newcommand{\Ham}{H}
\newcommand{\nop}{n}

\newcommand\ud[2]{\genfrac{}{}{0pt}{}{#1}{#2}}

\newcommand{\Zd}[1]{\mathbf{Z}_{#1}}

\begin{document}
\title{Quantum Many-Body Scars beyond the PXP model in Rydberg simulators}
\author{Aron Kerschbaumer}
\affiliation{Institute of Science and Technology Austria (ISTA), Am Campus 1, 3400 Klosterneuburg, Austria}
\author{Marko Ljubotina}
\affiliation{Institute of Science and Technology Austria (ISTA), Am Campus 1, 3400 Klosterneuburg, Austria}
\affiliation{Technical University of Munich, TUM School of Natural Sciences, Physics Department, 85748 Garching, Germany}
\affiliation{Munich Center for Quantum Science and Technology (MCQST), Schellingstr. 4, München 80799, Germany}
\author{Maksym Serbyn}
\affiliation{Institute of Science and Technology Austria (ISTA), Am Campus 1, 3400 Klosterneuburg, Austria}
\author{Jean-Yves Desaules}
\email{jean-yves.desaules@ist.ac.at}
\affiliation{Institute of Science and Technology Austria (ISTA), Am Campus 1, 3400 Klosterneuburg, Austria}
\date{\today}

\begin{abstract}
Persistent revivals recently observed in Rydberg atom simulators have challenged our understanding of thermalization and attracted much interest to the concept of quantum many-body scars (QMBSs). QMBSs are non-thermal highly excited eigenstates that coexist with typical eigenstates in the spectrum of many-body Hamiltonians, and have since been reported in multiple theoretical models, including the so-called PXP model, approximately realized by Rydberg simulators. At the same time, questions of how common QMBSs are and in what models they are physically realized remain open. In this Letter, we demonstrate that QMBSs exist in a broader family of models that includes and generalizes PXP to longer-range constraints and states with different periodicity. We show that in each model, \emph{multiple} QMBS families can be found. Each of them relies on a different approximate $\mathfrak{su}(2)$ algebra, leading to oscillatory dynamics in all cases. However, in contrast to the PXP model, their observation requires launching dynamics from weakly entangled initial states rather than from a product state. QMBSs reported here may be experimentally probed using Rydberg atom simulator in the regime of longer-range Rydberg blockades.
\end{abstract}

\maketitle

\noindent{\bf \em Introduction.---}%
Many-body quantum systems generically exhibit fast thermalization, where the system quickly loses memory of initial observables and equilibrates to the state determined only by conserved quantities such as energy. Such behavior can be understood~\cite{Alessio16} by assuming that eigenstates obey the eigenstate thermalization hypothesis (ETH)~\cite{SrednickiETH,Deutsch2018ETH}. The ETH conjectures that the eigenstates themselves behave like thermal states and naturally explains thermalization observed in unitary dynamics. In the examples where interacting quantum systems do not reach a simple thermal state, such as in so-called integrable~\cite{sutherland2004beautiful} and many-body localized models~\cite{Abanin2019Colloquium,HuseReview}, one typically observes that ETH is also violated by all eigenstates of the system. The strong breakdown of ETH is attributed to the emergence of an extensive number of conserved quantities, that constrain the dynamics and impact properties of eigenstates.

More recently, the phenomenon of quantum many-body scars (QMBSs)~\cite{Serbyn2021Review,Chandran2023Review,Moudgalya2022Review} demonstrated a weak breakdown of ETH. QMBSs are defined as eigenstates of interacting quantum systems that violate ETH, while coexisting with other ETH-complying eigenstates at finite energy density. Although such isolated non-thermal eigenstates were previously found in several models~\cite{Arovas1989AKLT,Shiraishi17,Moudgalya18}, the experimental observation of persistent revivals in a Rydberg atom quantum simulator~\cite{Bernien2017Rydberg} and its theoretical understanding through scars~\cite{TurnerNature,TurnerPRB} attracted significant attention.

One of the most emblematic models featuring QMBSs is the so-called PXP model~\cite{Lesanovsky2012,FendleySachdev}, that approximately describes the Rydberg atom experiment~\cite{Bernien2017Rydberg} in the nearest-neighbor blockade regime where scarring occurs. In this limit, each Rydberg atom -- viewed as a two-level system -- performs Rabi oscillations under a driving field, subject to a blockade condition that prevents adjacent atoms from simultaneously being in the excited state. Theoretically, QMBSs oscillations in the PXP model were linked to an approximate hidden $\mathfrak{su}(2)$ algebra~\cite{TurnerPRB,Choi2018su2} or, alternatively, to periodic trajectories in a variational description~\cite{Wenwei2018TDVP,Michailidis2020Mixed,TurnerPRX}. However, despite a vast amount of research on the PXP model~\cite{Surace2020Rydberg,IadecolaMagnons,Motrunich19,Rozon2022Floquet,Giudici2023Unraveling,Windt2022Squeezing}, its properties are still not fully understood. 

In particular, recent works hint at the existence of numerous other approximate QMBSs eigenstates~\cite{Kuba22,LjubotinaPRX} in the PXP model. Nevertheless, a simple criterion to determine which states in the PXP model evade thermalization remains missing. Another mystery is the relation between scarring and the nearest-neighbor blockade condition: while generalizations of the PXP model with higher spins~\cite{Wenwei2018TDVP,Desaules2023TSM,Desaules2023QLM} have been numerically shown to host the same type of QMBSs, this was not reported in generalizations of PXP that describe Rydberg arrays in regimes beyond nearest-neighbor blockade~\cite{ABlock}.

In this Letter, we demonstrate the existence of QMBSs in generalizations of the PXP model to \emph{arbitrary} blockade range. Constructing QMBSs in such generalized models and probing them experimentally, relies on initial \emph{short-range entangled} states, playing the role of the N\'eel product state used in the PXP model~\cite{Bernien2017Rydberg,TurnerNature,TurnerPRB}. For the PXP model and its generalizations, we construct three short-range parent states with different periodicity, that can be experimentally prepared, and lead to periodic revivals. Thus we broaden the class of physically realizable models, with each of them hosting at least three different families of QMBSs. As an implication of our results, we show that QMBSs-related dynamics can be used to efficiently prepare entangled $W$-states with a size of the order of the blockade range, and conjecture the existence of QMBS in the continuum limit. 

\noindent{\bf \em Models and properties.---}The PXP model~\cite{Lesanovsky2012,FendleySachdev}, 
$\Ham_\text{PXP}=\sum_{j=1}^N \PR_{j-1}\sxN_j\PR_{j+1},$
 is an approximate description of a $N$-atom Rydberg chain performing Rabi oscillations in the regime of nearest-neighbor blockade. Each Rydberg atom is viewed as an effective spin-$1/2$ either in the ground ($\ket{\da}$) or excited Rydberg ($\ket{\ua}$) state, and the Pauli matrix $\sigma^x_i$ generates Rabi oscillations. Since atoms interact only in their Rydberg ($\ket{\ua}$) states, tuning the inter-atom distance can prevent nearest-neighbor excitations via the Rydberg blockade. In the PXP Hamiltonian, this condition is implemented using the projector $\PR=\ket{\da}\bra{\da}$. 

While the PXP Hamiltonian is non-integrable~\cite{TurnerNature,Khe19,Park2024,Balazs2024}, it features a family of QMBSs related to the N\'eel state $\ket{\Zd{2}}=\ket{\da\ua\da\ua \ldots \da\ua}$~\cite{ZDconv}. QMBSs are approximately contained in the subspace that can be generated by the action of a simple raising operator onto the N\'eel state -- a manifestation of an approximate $\mathfrak{su}(2)$ algebra~\cite{Choi2018su2,Bull2020Lie}. Consequently, unitary dynamics launched from the N\'eel state leads to coherent oscillations between $\ket{\Zd{2}}$ and $\ket{\Zd{2}'}=\ket{\ua\da\ua\da \ldots \ua\da}$. Revivals can also be seen from the period-3 state $\ket{\Zd{3}}=\ket{\da\da\ua\da\da\ua \ldots \da\da\ua}$. However, $\ket{\Zd{n}}$ states with $n>3$ equilibrate quickly despite their similar structure.

We generalize the PXP model to blockade radius $\alpha$ as follows:
\begin{equation}\label{Eq:PXPa}
    H_\alpha{=}\sum_{j=1}^N \PR_{j-\alpha}\ldots  \PR_{j-1}\sigma^x_j  \PR_{j+1}\ldots\PR_{j+\alpha}{=}\sum_{j=1}^N \sx_j ,
\end{equation}
which can be approximately realized experimentally, see App.~\ref{app:exp}. This model is the focus of this Letter and reduces to $\Ham_\text{PXP}$ when $\alpha=1$.
Here and in the rest of this Letter, we use $\sx_j$ as a shorthand notation for $\sxN_j$ dressed with $\alpha$ projectors on each side. We also assume periodic boundary conditions (PBC) unless stated otherwise. 
For brevity, we generally suppress $\alpha$ except in $H_\alpha$.

\begin{figure}[tb]
\centering
\includegraphics[width=0.9\columnwidth]{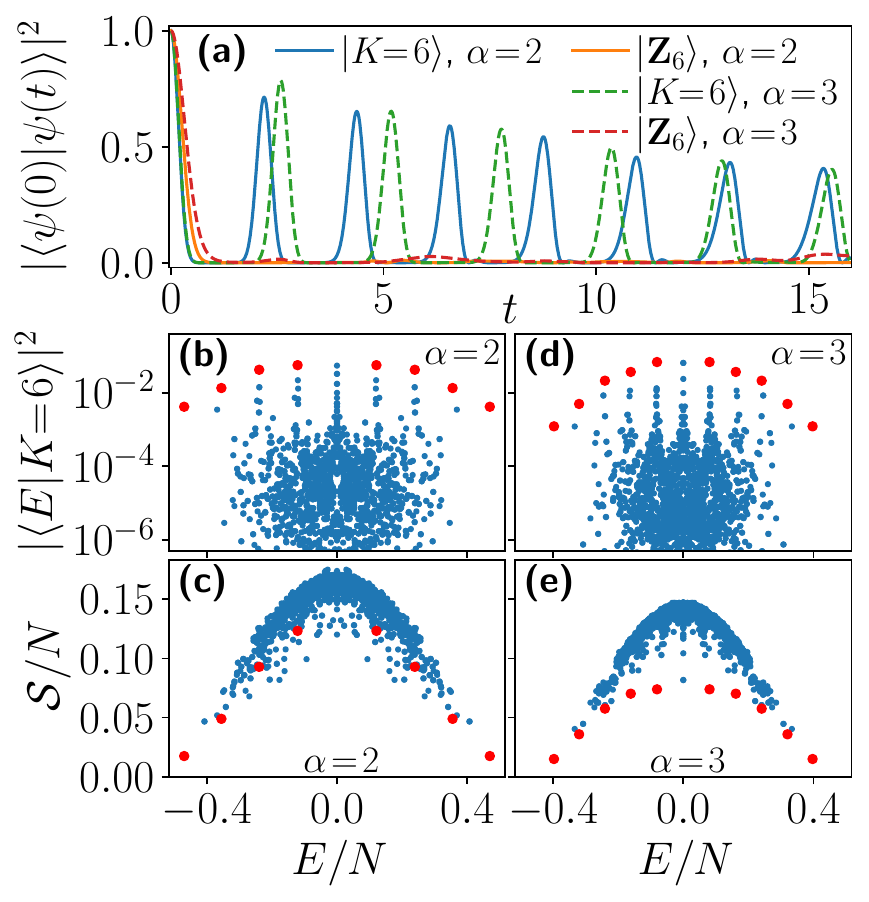}
\caption{QMBSs with longer-range blockade $\alpha{=}2$ and 3 obtained from the $\ket{K{=}6}$ state. 
(a) Time evolution after a quench shows good revivals for the $\ket{K{=}6}$ state and no revivals for the product state $\ket{\Zd{6}}$. Overlap of eigenstates with $\ket{K{=}6}$ in (b) and (d) as well as half-chain von Neumann entanglement entropy $S$ of the eigenstates in (c) and (e) show QMBSs as outliers, that are highlighted in red and coincide in plots (b)-(c) and (d)-(e). Data shown is for system sizes $N{=}24,\,30$ (for $\alpha{=}2,\, 3$ respectively) and all relevant momentum sectors.
}
\label{fig:olap_v2}%
\end{figure}

The Hamiltonian~(\ref{Eq:PXPa}) is chaotic for generic $\alpha$ and a few isolated exact scars are known~\cite{Surace2021Exact, Balazs2024}. However, these scarred eigenstates do not admit an algebraic description and no simple reviving state was reported for $\alpha>1$. Even $\ket{\Zd{\alpha+1}}$, the natural generalization of $\ket{\Zd{2}}$ to $\alpha>1$, displays fast thermalization under dynamics generated by $H_\alpha$. Intuitively, this absence of revivals can be attributed to \emph{frustration}. For $\alpha=1$ the N\'eel state $\ket{\Zd{2}}$ has a unique partner $\ket{\Zd{2}^\prime}$ under translation, leading to oscillations between these two states. However, for $\alpha\geq 2$, $\ket{\Zd{\alpha + 1}}$ has $\alpha$ translated partners. A superposition of these states cannot serve as the unique partner state for $\ket{\Zd{\alpha + 1}}$ since finite-time dynamics is incapable of transforming a given product state into a cat-like superposition of macroscopically different states, resulting in frustration. Below, we argue that introducing short-range entanglement in the initial state lifts this frustration~\cite{Cirac21,Nielsen04,Preskill} as it allows to create new pairs of states between which scarred dynamics is possible.

\noindent{\bf \em Lifting frustration by entanglement.---}%
We propose a different way of generalizing the N\'eel state for an even unit-cell size $K$ (we discuss the odd case below). We define the state $\ket{K}$ with $M$ repeating unit cells of size $K$ as 
\begin{equation}\label{Eq:K-def}
\ket{K}=\bigotimes_{j=1}^M \left[\ket{0}_{K/2}\otimes \ket{W}_{K/2} \right], 
\end{equation}
that is built from the product state $\ket{0}_{K/2}=\ket{\da\ldots \da}$, in the first half of the unit cell, and the entangled $W$-state, $\ket{W}_{K/2}=\sqrt{2/K}(\ket{\ua\da\cdots \da}+\ket{\da\ua\da\cdots \da}+\ldots+\ket{\da\cdots \da\ua})$ where one $\ua$-spin is uniformly distributed in the second half of the unit cell. This short-range entangled state can be written as a matrix product state (MPS) with bond dimension $\chi=2$ for $K>2$ and reduces to the N\'eel state for $K=2$. Below we demonstrate that this state lifts frustration, such that it is approximately transformed into its reflected version $\ket{K^\prime}=\sr \ket{K}$ (with $\sr$ the spatial reflection operator) with unit cell $\ket{W}_{K/2}\otimes\ket{0}_{K/2}$ under the unitary dynamics generated by $H_\alpha$.

Figure~\ref{fig:olap_v2}(a) shows that for a period $K=6$ unit cell, the $\ket{K}$ state has strong revivals in models with $\alpha=2$ and $\alpha=3$. These revivals are indicative of the presence of QMBSs~\cite{Alhambra2020}.
QMBSs become apparent in Fig.~\ref{fig:olap_v2}(b)-(e) which show the overlap of eigenstates of $H_{\alpha=2,3}$ with the $\ket{K}$ state as well as their entanglement entropy. A set of approximately equally spaced eigenstates is clearly visible with enhanced overlap and low entanglement. 

For a chain of length $N$, QMBSs appear at $2N/K+1$ special energies, i.e. twice the number of unit cells plus one. However, there are more scarred eigenstates as some of them are degenerate, see SM~\cite{SM}.
This number of scarred energies can be understood by considering every cell as an effective two-level system oscillating between $\ket{0}_{K/2}\otimes \ket{W}_{K/2}\equiv \ket{\Downarrow \Uparrow}$ and $\ket{W}_{K/2}\otimes\ket{0}_{K/2}\equiv \ket{\Uparrow \Downarrow}$, providing an approximate mapping to the PXP $\ket{\Zd{2}}$ case with $\ket{0}\rightleftharpoons \ket{\Downarrow}$ and $\ket{W}\rightleftharpoons \ket{\Uparrow}$. For such a mapping to work, the Rydberg blockade must be effective within the range of $K/2$: the application of $H_\alpha$ to the state $\ket{W}_{K/2}$ should yield the $\ket{0}_{K/2}$ state and not be able to create additional $\ket{\uparrow}$-excitations. This leads to an upper bound on the unit cell size, $K/2\leq \alpha+1$. In addition, we obtain the lower bound $K\geq2\alpha$, since values of $K$ below $2\alpha$ cause excitations in adjacent $\ket{W}_{K/2}$ states to violate the blockade condition. This leaves only \emph{three} possible values of the unit cell size $K$ for a given $\alpha$, $K=2\alpha$, $K=2\alpha+1$, and $K=2\alpha+2$. We first consider even $K$ and later generalize to $K=2\alpha+1$.

\noindent{\bf \em Algebraic description.---}%
To devise the $\mathfrak{su}(2)$ algebra, we generalize the forward scattering approximation (FSA) introduced for the PXP model~\cite{TurnerNature}. We set the global raising operator $\Jp$ such that $\Jp+\Jm=\Ham_\alpha$ and $\Jm\ket{K}=0$, where $\Jm\coloneqq\left(\Jp\right)^\dagger$. This implies that $\Jp$ acts with $\spp$ on sites within $\ket{0}_{K/2}$ in Eq.~(\ref{Eq:K-def}) and as $\smm$ on sites within $\ket{W}_{K/2}$. These operators are related by inversion, $\Jm=\sr\Jp \sr$.
They allow us to define the rest of the algebra, with 
$\Jx=\left(\Jp{+}\Jm\right)/2 = H_\alpha/2$ and 
\begin{equation}\label{Eq:Jy}
\Jy=\frac{1}{2i}\left(\Jp{-}\Jm\right) =\frac{1}{2} \sum_{j=1}^Nf(j)\sy_j,
\end{equation}
being the sum of $\sy$ operators with positive (negative) $f(j)=\pm1$ signs on the first (last) $K/2$ sites of each unit cell.

We obtain the $z$-component of the collective spin as $J^z = [\Jx,\Jy]/i$. It has $\sz$ terms with the same $f(j)$ sign structure as in $J^y$. This favors $\ket{\da}$ in the first half of each cell and $\ket{\ua}$ in the second half.
Unlike in the PXP $\Zd{2}$ case, $\Jz$ also has $XY$-type terms $f(j)(\spp_j\smm_k+\spp_k\smm_j)$. Crucially, for $K=2\alpha$ or $2\alpha+2$ these terms act between \emph{all} sites in the second half of the cell, favoring $\ket{W}$ over any other combination of states with a single $\ket{\ua}$. As a consequence, the ground state $\GSz$ of $\Jz$ is approximately equal to $\ket{K}$ ($\approx 0.98$ overlap for the cases of $\alpha$ and $N$ shown in Fig.~\ref{fig:olap_v2}), thus providing a self-consistent check of this algebra. Meanwhile, for other (even) values of $K$, the $XY$ terms have a different structure and the overlap between $\GSz$ and $\ket{K}$ drops quickly.

\begin{figure}[t]
\centering
\includegraphics[width=\columnwidth]{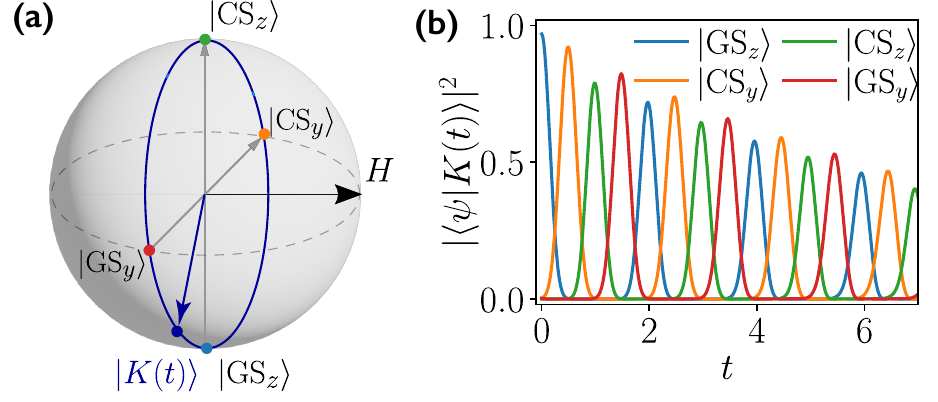}
\caption{ (a) Collective-spin picture of the dynamics. Starting from $\ket{K}\approx \GSz$, we expect the Hamiltonian (proportional to $\Jx$) to generate precession in the $yz$-plane, passing through the ground- and ceiling-states of $\Jy$ and $\Jz$. This dynamics is illustrated by plotting fidelities for $\alpha=4$, $K=9$, and $N=36$ in panel (b).}
\label{fig:algebra}%
\end{figure}

From the algebraic picture, we can understand the revivals as the precession of a collective spin of size $N/K$. Starting from $\ket{K}\approx \GSz$, the Hamiltonian -- which is proportional to $\Jx$ -- will lead to a rotation in the $yz$-plane along the Bloch sphere. We then expect the wavefunction to pass close to the ceiling state $\CSy$ of $\Jy$, the ceiling state $\CSz=\sr \GSz \approx \ket{K^\prime}$ of $\Jz$, the ground state $\GSy=\sr\CSy$ of $\Jy$, before coming back to $\GSz$, as indicated in Fig.~\ref{fig:algebra}(a).

Our Ansatz for the algebra can be extended to odd values of $K$ by acting with $\sx/2$ (instead of $\spm$) on the middle site of each unit cell in $\Jp$. This preserves the property $\sr \Jp \sr=\Jm \coloneqq (\Jp)^\dagger$. The remaining operators of the approximate $\mathfrak{su}(2)$ algebra are constructed as for even $K$ above. We leave the details to Appendix~\ref{app:algebra} and focus on the ground state of $\Jz$ which is approximately 
\begin{equation}\label{Eq:Kodd}
\ket{K}=\bigotimes_{j=1}^M \left[\ket{0}_{\alpha}\otimes \ket{\tilde W_\beta}_{\alpha+1} \right],
\end{equation}
for $K=2\alpha+1$. The deformed $W$-state 
$
    \ket{\tilde W_\beta}_{\alpha+1}=[\beta\ket{\ua}\ket{0}_\alpha{+}\sqrt{\alpha}\ket{\da}\ket{W}_\alpha]/{\sqrt{|\beta|^2{+}\alpha}}
$
(with a numerically determined optimal value of $\beta\approx0.65$), also contains superpositions of all possible one-spin excitations within the block of $\alpha+1$ sites, but the weight of excitation on the first spin is now suppressed. 
Figure~\ref{fig:algebra}(b) shows that a quench from state~(\ref{Eq:Kodd}) leads to strong oscillatory dynamics, as was the case for $K=2\alpha$ and $K=2\alpha+2$. Furthermore, the overlap with the ground and ceiling states of $\Jy$ and $\Jz$ demonstrates that the picture of precession of a large spin offered by the approximate $\mathfrak{su}(2)$ algebra holds well. 

Inspired by this description, we develop an MPS variational manifold akin to spin coherent states projected in the constrained Hilbert space~\cite{Wenwei2018TDVP,Michailidis2020Mixed}. In Appendix~\ref{app:TDVP}, we present evidence of the existence of periodic trajectories in this semiclassical limit for $\alpha=1$ and $\alpha=2$, with a period $K=4$ unit cell using MPS with bond dimension $\chi=2$ and $\chi=3$ respectively. We conjecture that such trajectories exist for any blockade range. 

\noindent{\bf \em Implications for scars in PXP.---}The $\ket{K}$ states constructed in Eqs.~(\ref{Eq:K-def}) and (\ref{Eq:Kodd}) also have non-trivial implications for QMBSs in the PXP model, $\alpha=1$. 
In that model, a weakly entangled period-3 initial state was known to have revivals better than $\ket{\Zd{3}}$~\cite{Michailidis2020Mixed}. Our approach elucidates its origin, as we find that it corresponds to the ground state of $\Jy$.
In contrast, revivals from period-4 states in the PXP model were not reported previously. This is explained by the small overlap of $\ket{\Zd{4}}$ on the proper reviving state built from the size-4 unit cell $\ket{\da\da}[\ket{\ua\da}+\ket{\ua\da}]/\sqrt2$ [cf.~with Eq.~(\ref{Eq:K-def}) for $\alpha=1$ and $K=4$]. For instance, for $N=24$ the overlap of $\ket{\Zd{4}}$ with the period-4 state Eq.~(\ref{Eq:K-def}) is very small $0.841^N\approx 0.0156$, whereas $\ket{\Zd{3}}$ has considerable overlap with the best reviving period-3 state at $0.956^N \approx 0.336$~\cite{SM}.

\begin{figure}[t]
\centering
\includegraphics[width=\columnwidth]{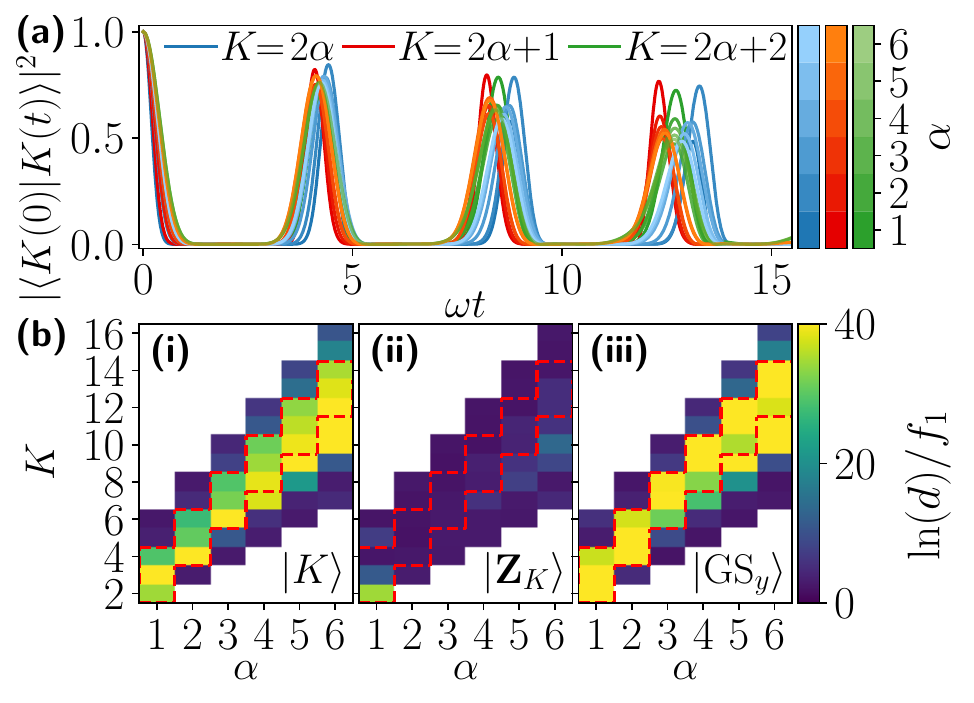}
\caption{(a) Fidelity after quenches from the $\ket{K}$ state for $2\alpha\leq K\leq 2\alpha+2$. Time is rescaled by $\omega=\sqrt{K/2+C}$, with $C=-0.15$ for $K\leq 2\alpha+1$ and $C=0.75$ for $K=2\alpha+2$. (b) Inverse fidelity density $f_1$ times the logarithm of the quantum dimension $d$ for three different initial states. A large value indicates clear revivals. The red dashed lines denote the range between $K=2\alpha$ and $K=2\alpha+2$. In the domain they delimit, good revivals can be seen for all $\alpha$ for $\ket{K}$ and $\GSy$, but not for $\ket{\Zd{K}}$. The system sizes used and the data for fidelity can be found in SM~\cite{SM}.
}
\label{fig:all_revs}
\end{figure}

\noindent{\bf \em Revivals across different blockade ranges.---}%
We compare dynamics for the reviving states found above across different $\alpha$. Figure~\ref{fig:all_revs} shows clear fidelity revivals after quenches from the $\ket{K}$ state for $\alpha=1,\ldots, 6$ and the three special unit cell sizes discussed above. Importantly, the revival period scales approximately as $1/\sqrt{K/2+C}$. This can be understood from the approximate mapping to the PXP $\ket{\Zd{2}}$ case using an effective two-level system. Mapping $\ket{W}_{K/2}$ to $\ket{\Uparrow}$ and $\ket{0}_{K/2}$ to $\ket{\Downarrow}$ gives the $\ket{\Zd{2}}$ state of effective spins, but the matrix element of the Hamiltonian is now $\bra{\Uparrow}H_\alpha\ket{\Downarrow}=\sqrt{K/2}$. This implies a scaling of the revival period as $\sqrt{2/K}$ without any corrections. However, this mapping is only approximate, and numerically we find that $C\approx -0.15$ for $K=2\alpha$ and $2\alpha+1$, and $C\approx 0.75$ for $K=2\alpha+2$.

Next, we compare revivals across a much broader range of unit cell sizes, $\alpha+1 \leq K\leq 2\alpha+4$ for different $\alpha$. Note that the initial state $\ket{K}$ violates the blockade condition for $K<2\alpha$, hence we project it back onto the constrained subspace. Revivals can be quantified in a system-size independent way using the fidelity density $f_1$ as this quantity quickly converges with $N$~\cite{SM}. Fidelity density is defined as $f_1=-\ln( \mathcal{F}_1)/N$, where $\mathcal{F}_1 = |\bra{K} e^{-iTH_\alpha}\ket{K}|^2$ is the fidelity at the time of the first revival. However, we also have to account for the varying constraint as $\alpha$ is changed, which is achieved by using a normalized inverse fidelity density, $\ln(d)/f_1$, with $d$ the effective local Hilbert space dimension (such that the Hilbert size is $\sim d^N$~\cite{SM}). The quantity $\ln (d)/f_1$ is expected to be of order one for a thermalizing state, and much larger than one in the case of strong revivals, indicating ergodicity breaking. For large $\alpha$, Fig.~\ref{fig:all_revs}~(b) shows revivals not only for the three special values of $K$, but also for $K<2\alpha$. In contrast, for 
$K>2\alpha+2$ the revivals disappear rapidly. We conjecture that the revivals for $K<2\alpha$ can be understood in a similar fashion, but with initial states presenting longer-range correlations.

Figure~\ref{fig:all_revs}(b) also shows that initial product states do not have clear revivals except for $\ket{\Zd{2}}$ in the PXP model, whereas the initial $\GSy$ state obtained from $\mathfrak{su}(2)$ algebra has revivals that are stronger compared to the $\ket{K}$ initial state  (in particular for $K=2\alpha+1$, see SM~\cite{SM}). Below we use the good revivals of $\GSy$ to devise an experimental protocol for probing QMBSs.

\begin{figure}[b]
\centering
\includegraphics[width=0.98\columnwidth]{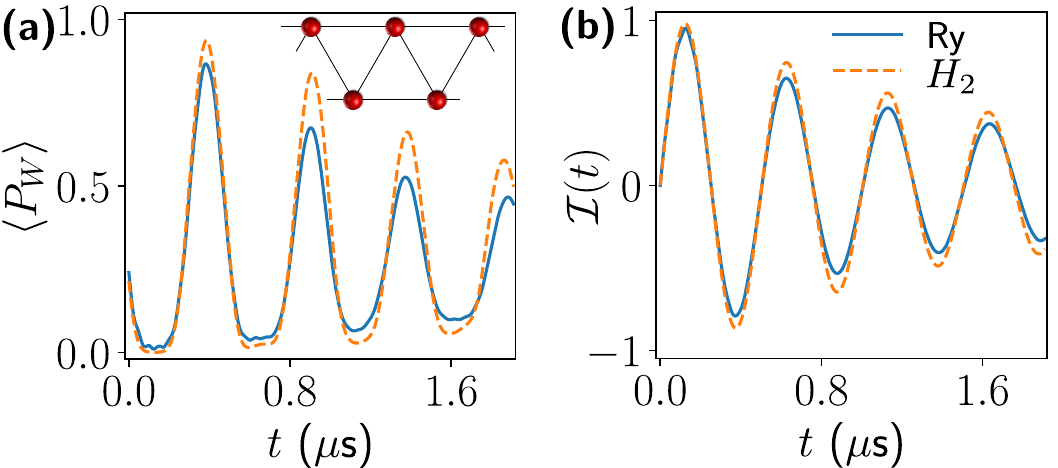}
\caption{
 (a) Expectation value of the projector onto a local $W$-state and imbalance (b) show long time coherent dynamics, with only a small difference between the approximate $H_2$ Hamiltonian with (dashed lines) and the realistic Ising-type Hamiltonian of Rydberg atoms arranged into a triangular ladder (see inset and App.~\ref{app:exp}) with $N=20$ sites and OBC. 
} 
\label{fig:fig4}
\end{figure}

\noindent{\bf \em Experimental implementation.---}%
We focus on the Hamiltonian $H_2$ from Eq.~(\ref{Eq:PXPa}) that can be approximately implemented using a triangular ladder of Rydberg atoms, see inset in Fig.~\ref{fig:fig4}(a). Next, we propose to prepare the ground state of the $\Jy$ operator in Eq.~(\ref{Eq:Jy}), $\GSy$, instead of $\ket{K}$ from Eq.~(\ref{Eq:K-def}). This state is also a point on the trajectory of the dynamics launched from the $\ket{K}$ state, see Fig.~\ref{fig:algebra}(b), and it can be naturally prepared from the ground state of $H_2$ (obtainable via annealing~\cite{Bernien2017Rydberg,Keesling19}) by single-site unitary rotations ($\pm \pi/2$ pulses in the $z$-direction).

The time evolution with $H_2$ from the $\GSy$ state is expected to give long-time coherent oscillations, approximately passing different points on the collective spin orbit in Fig.~\ref{fig:algebra}(b). After a quarter-period, the unitary dynamics will approximately lead to the $\ket{K}$ state, which in this case consists of $W$-states of range two interlaced by product states. Figure~\ref{fig:fig4}(a) confirms these expectations by showing the expectation value of the projector onto a two-site $W$-state, $P_W = (\ket{\ua\da}{+}\ket{\da\ua})(\bra{\ua\da}{+}\bra{\da\ua})/2$ that reaches a value close to one after a quarter period. The dynamics of the average local imbalance $\mathcal{I}=({4}/{N})\sum_{j=0}^{N/4-1} [\langle \nop_{4j+3}\rangle{+}\langle \nop_{4j+4}\rangle{-}\langle \nop_{4j+1}\rangle{-}\langle \nop_{4j+2}\rangle]$ shown in Fig.~\ref{fig:fig4}(b) also features coherent oscillations. Notably, the dynamics generated by the approximate $H_2$ Hamiltonian and by the Ising Hamiltonian for Rydberg atoms arranged into a triangular ladder, where the range-two Rydberg blockade effectively emerges, are similar.

\noindent{\bf \em Conclusion and outlook.---}%
We demonstrated the existence of new families of QMBSs in longer-range generalizations of the PXP model up to a blockade radius of six. This leads to a natural conjecture that this phenomenon persists for an arbitrary blockade radius. We constructed an effective description of the dynamics in terms of a collective $\mathfrak{su}(2)$ spin, and argued that in order to lift frustration, the QMBSs and non-ergodic dynamics must originate from weakly entangled initial states. We also confirmed the existence of ``semiclassical'' periodic trajectories in variational dynamics over MPS~\cite{Wenwei2018TDVP,Michailidis2020Mixed}, thus
contributing to the ongoing discussion on the similarities between QMBSs in PXP-type models and scars in models with a more physical classical limit, both in single-\cite{Heller84} and many-body~\cite{pilatowsky-cameo_ubiquitous_2021, evrard_quantum_2024, hummel_genuine_2023, pizzi_quantum_2024, ermakov_periodic_2024} quantum systems.

Our work opens the door to the study of QMBSs and weak ergodicity breaking in Rydberg models with longer-range blockades, and of potential stabilization via deformations~\cite{Choi2018su2,Khe19,Bull2020Lie,Omiya2022} or driving~\cite{Bluvstein2021Controlling,Maskara21}. Although ground state properties were studied for these models~\cite{Keesling19,Sachdev20,Che21,Che21-2,Che22}, excited state dynamics received little attention. More importantly, our results suggest that QMBSs may be more commonly related to initial weakly entangled states, in agreement with the understanding of QMBSs as originating from periodic trajectories that generally do not pass through product states~\cite{Wenwei2018TDVP,Michailidis2020Mixed}. Moreover, our generalization of QMBSs to a family of models with an arbitrary blockade range suggests a possible extension to continuous (non-lattice) model with QMBSs.

\noindent{\bf \em Acknowledgments.---}%
The authors are grateful to Zlatko Papi\'c, Dolev Bluvstein, Nishad Maskara, Marcello Dalmonte, Thomas Iadecola and Johannes Feldmeier for insightful discussions. 
A.~K., M.~L. and M.~S. acknowledge support by the European Research Council under the European Union's Horizon 2020 research and innovation program (Grant Agreement No.~850899).
J.-Y.D.~acknowledges funding from the European Union’s Horizon 2020 research and innovation programme under the Marie Sk\l odowska-Curie Grant Agreement No.~101034413.

\noindent{\bf \em Data availability.---}%
The data presented in this Letter is available at \cite{data_ISTAREX}.

\bibliography{biblio}

\begin{thebibliography}{61}
\bibitem [{ZDc()}]{ZDconv}%
  \BibitemOpen
  \href@noop {} {}\bibinfo {note} {In this work, we will use the convention for
  all $\Zd{d}$ type states that the up-spin is located at the rightmost site of
  the unit-cell.}\BibitemShut {Stop}%
\end{thebibliography}

\clearpage 
\pagebreak
\begin{appendix}

\section{Experimental setup}\label{app:exp}
The experimental Hamiltonian we consider is 
\begin{equation}\label{eq:H_exp}
    \frac{1}{\hbar}\Ham_\mathrm{exp}=\frac{\Omega}{2}\sum_{j=1}^N \sxN_j -\Delta\sum_{j=1}^N \nop_j+V_1\sum_{\substack{j,k=1 \\ j<k}}^N \frac{\nop_j\nop_k}{d_{j,k}^6},
    \vspace{-2mm}
\end{equation}
where $d_{j,k}=||\vec{r}_j-\vec{r}_k||_2/a$ is the distance between atoms $j$ and $k$ in units of the lattice spacing $a$. The Rydberg blockade prevents excitation within the blockade distance $R_b$, defined by $\Omega = V_1 (a/R_b)^6$~\cite{Bernien2017Rydberg}. Choosing the lattice spacing  $R_b / (\alpha + 1) < a < R_b / \alpha$  then approximates the Hamiltonian in Eq.~(\ref{Eq:PXPa}). However, for large  $\alpha$, maintaining a strong blockade while suppressing long-range interactions becomes increasingly challenging. For the case $\alpha=2$, these problems can be mitigated. Instead of a linear chain, we consider a triangular ladder such that each atom (away from the boundaries) has exactly 4 nearest-neighbors with $d=1$. This allows us to more efficiently implement the PPXPP model ($\alpha=2$). We set $\Omega/2\pi=2$ MHz and use $V_1/\Omega=9$. We also use $\Delta/\Omega=0.213$ instead of 0 to further suppress long-range interactions, see SM of Ref.~\cite{Bluvstein2021Controlling}. The Hamiltonian in Eq.~(\ref{eq:H_exp}) was used to produce the data in Fig.~\ref{fig:fig4} using sparse Krylov methods.

\vspace{-4mm}
\section{Algebraic structure}\label{app:algebra}
For even values of $K$  and for any $\alpha$, we define the FSA raising operator $\Jp$ as
\begin{equation}
\Jp=\sum_{j=0}^{N/K-1}\Big[\sum_{l=1}^{K/2} \spp_{Kj+l} +\sum_{l=K/2+1}^{K} \smm_{Kj+l}\Big].
\end{equation}
This operator takes $\ket{0}_{K/2}$ to $\ket{W}_{K/2}$ in the first half of the cell and does the opposite in the second half. The lowering operator is $\Jm=\left(\Jp\right)^\dagger$. We can use these two operators to define $\Jx=\left(\Jp{+}\Jm\right)/2=\Ham_\alpha/2$. We can also define $ \Jy$ as in Eq.~(\ref{Eq:Jy}) of the main text. 
We compute the $\Jz$ operator as $[\Jx,\Jy]/i$. Due to the form of $\Jx$ and $\Jy$, $\Jz$ will only have diagonal terms of the form $\sz$ as well as off-diagonal terms $\spp_l\smm_{m}+\smm_l\spp_{m}$. In general, these terms go from range-1, $\spp_l\smm_{l+1}+\smm_l\spp_{l+1}$, to range-$\alpha$, $\spp_l\smm_{l+\alpha}+\smm_l\spp_{l+\alpha}$. However, there can be cancellations between them for specific values of $K$.
Indeed, if $K=2\alpha$ or $K=2\alpha+2$, the two halves of the cells are essentially disconnected and $\Jz$ simplifies to
\begin{align}
    &\Jz{=}\frac{1}{2}
    \sum_{j=0}^{N/K-1}\Bigg[
    \Big[ \sum_{l=1}^{K/2} -\sum_{l=k+1}^{K}\Big]\sz_{Kj+l}\\
  &+ 
  \Big[
  \sum_{\substack{l,m=1 \\ l<m}}^{K/2}-\sum_{\substack{l,m=K/2+1 \\ l<m}}^{K}
  \Big]
    (\spp_{Kj+l}\smm_{Kj+m}{+}\spp_{Kj+m}\smm_{Kj+l})\Bigg] \nonumber
\end{align}
The $XY$-type terms act between all possible pairs of sites in the first half of the cell (with a $+1$ prefactor) and between all possibles pairs of sites in the second half of the cell (with a $-1$ prefactor). In SM~\cite{SM} we show details for a few values of $\alpha$ and $K$, including cases where $K$ is not between $2\alpha$ and $2\alpha+2$.  

For $K$ odd, the $\Jp$ operator has a similar structure, with $\spp$ in the first half of the cell (first $(K-1)/2$ spins) and $\smm$ in the second half, but the middle site now has an $\sx/2$ operator. This preserves the fact that $\Jm$ can be obtained by applying spatial inversion to $\Jp$, and that $\Jp$ and $\Jm$ still sum to $\Ham$. This implies that the $J^y$ operator is as in the even $K$ case (with $\sy$ in the first half of the cell and $-\sy$ in the second half) but without any operators acting on the middle site of the cell.

In the special case where $K=2\alpha+1$, the $\Jz$ operator has a similar structure as for $K=2\alpha$ and $K=2\alpha+2$. The same $XY$ terms act in the first half of each cell (sites 1 to $\alpha$) as in the second half (sites $\alpha+2$ to $2\alpha+1$). In addition, there are also $XY$ terms between both halves of the cell and the middle site $\alpha+1$. These additional terms in $\Jz$ are formally written as
\begin{equation}
    \frac{1}{4}\big[\sum_{l=1}^{\alpha} -\hspace{-0.22cm} \sum_{l=\alpha+2}^{K}\big]
\hspace{-0.08cm}\left(\spp_{Kj+l}\smm_{Kj+\alpha+1}{+}\spp_{Kj+\alpha+1}\smm_{Kj+l}\right),
\end{equation}
where we omitted the sum over unit cells that runs from $j=0$ to $j=N/K-1$.

\vspace{-3mm}
\section{Periodic MPS trajectories}\label{app:TDVP}
A low dimensional manifold which can capture the scarring from the $\ket{\Zd{2}}$ and $\ket{\Zd{3}}$ states has previously been proposed for the PXP model~\cite{Wenwei2018TDVP,Michailidis2020Mixed}. 
It is based on parameterizing the wavefunction using a bond dimension $\chi=2$ matrix-product state (MPS) Ansatz
\begin{equation}
    A_j=\begin{pmatrix}
\cos(\theta_j)|{\downarrow}\rangle & e^{i\phi_j} \sin(\theta_j)|{\uparrow}\rangle \\
|{\downarrow}\rangle & 0 \\
\end{pmatrix}.
\end{equation}
This Ansatz can be thought of as a product of single-spin states on the Bloch sphere  $\cos(\theta) \ket{\da}+e^{i\phi}\sin(\theta)\ket{\ua}$, which is then projected into the constrained PXP space that allows no $\ket{\ldots\ua\ua\ldots}$ configurations. 
We can extend this Ansatz to larger blockade radii $\alpha$, with the bond dimension scaling as $\alpha+1$. 
This extended Ansatz reads
\begin{align}
A_j^{\alpha= 2}&=\begin{pmatrix}
\cos(\theta_j)|{\downarrow}\rangle & 0 & e^{i\phi_j}\sin(\theta_j)|{\uparrow}\rangle \\
|{\downarrow}\rangle & 0 & 0 \\
0 & |{\downarrow}\rangle & 0
\end{pmatrix},
\label{eq:mps-ansatz}
\end{align}
and analogously for longer-range blockades. For all $\alpha$ and $K$ satisfying $2\alpha\leq K \leq 2\alpha+2$, we find states in the manifold that have an overlap of over $97.8\%$ with $\GSy$ and over $94.2\%$ with $\GSz$ (for the system sizes used in Fig.~\ref{fig:all_revs}(b), see SM~\cite{SM} for details). This last number rises to over $98.3\%$ if we exclude the case $K=2\alpha+1$, highlighting the fact that the $J^z$ operator we constructed for odd $K$ is less accurate (see SM~\cite{SM} for more details). 

Crucially, we can always \emph{exactly} represent the $\ket{K}$ states for $2\alpha \leq K\leq 2\alpha+2$ from Eqs.~(\ref{Eq:K-def}) and (\ref{Eq:Kodd}) as MPS states using the above Ansatz for $A^\alpha_j$ with $K$ different tensors $A^\alpha_j$ forming the unit cell. These tensors are completely specified by the values of the two angles $(\theta_j,\phi_j)$.
For the even $K$ (with $K=2\alpha$ or $K=2\alpha+2$), the $\theta$ angles read: 
\begin{equation*}
    \theta_j{=}\begin{cases}
    0  &\text{if} \ j\leq K/2, \\
    a_{K+1-j }   &\text{if} \ j>K/2,
    \end{cases} \ \text{with}  \ \ud{a_1{=}\pi/2,}{ a_k{=}\arctan(\sin(a_{k-1})).}
\end{equation*} 
Meanwhile, all $\phi_j$ can be set to 0 as there is no phase difference depending on where the $\ket{\ua}$ are placed. The state $\ket{K=2\alpha+1}=\bigotimes_{j=1}^M [\ket{0}_{\alpha}\otimes \ket{\tilde W_\beta}_{\alpha+1}]$ is also contained in the MPS manifold. The first $\alpha$ angles and the last $\alpha$ angles have the same value as for the case $K=2\alpha$. For the middle site of the unit cell, the angle $\theta_{\alpha+1}$ is set to $\arctan(\beta\sin( \theta_{\alpha+2}))$, with the corresponding $\phi_{\alpha+1}$ angle being set to zero as well (assuming $\beta$ is positive). We note however that the $\ket{K}$ state can also be compressed to bond dimension $\chi=2$ for generic $K\ge2$ and $\chi=1$ for $K=2$. 

The projection of the unitary dynamics generated by $H_\alpha$ onto the MPS variational manifold yields first order differential equations. Following the scheme in Ref.~\cite{Michailidis2020Mixed}, equations of motion (EOMs) for the $\theta$ angles can be obtained from $\sum_j 2{\rm Im}\langle\partial_{\theta_j}\psi|\partial_{\phi_k}\psi\rangle\dot{\theta}_j=\partial_{\phi_k}\langle\psi|H_\alpha|\psi\rangle$ for all $k\in[1,K]$, where $\ket{\psi}$ is the MPS state with a unit cell of size $K$, depending on $K$ variational $\theta$ and $\phi$ angles. The values of the $\phi$ angles are set to $\phi_j=\pi/2$ after differentiation is performed, since the plane with $\phi_j=\pi/2$ forms a flow-invariant subspace where $\dot\phi_j=0 \ \forall j$.

\begin{figure}[t]
\centering
\includegraphics[width=0.85\columnwidth]{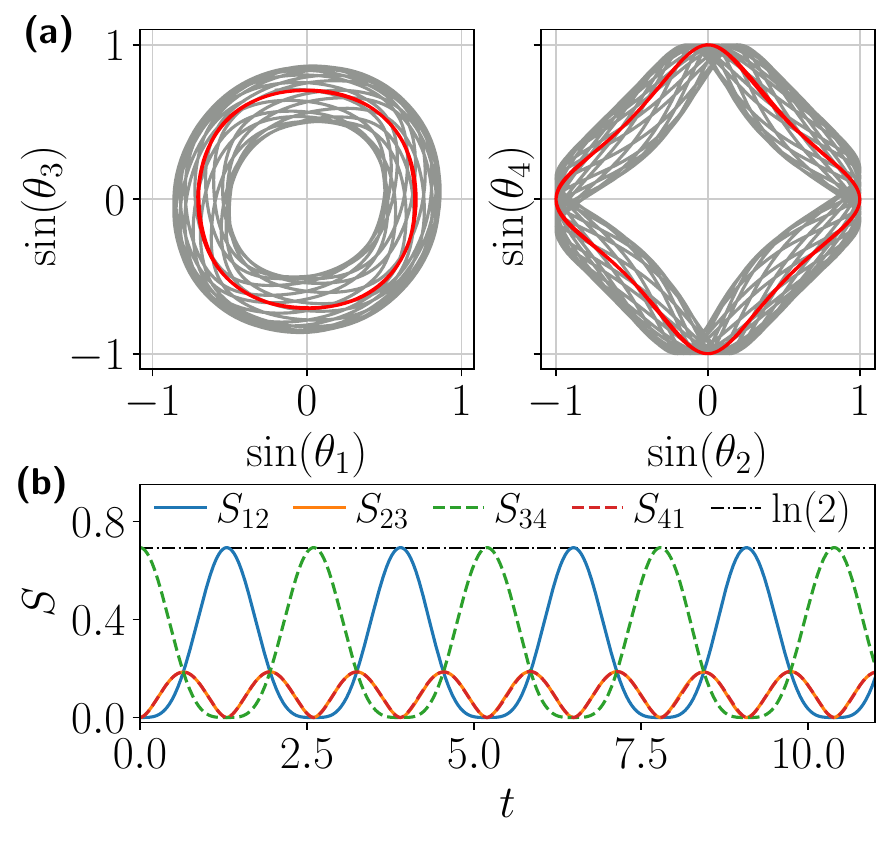}
\vspace{-4mm}
\caption{ (a) Dynamics in the space of the four angles parameterizing the MPS state after a quench from the $K$ state shows a closed periodic trajectory (red). Dynamics initialized from a perturbed initial point near the periodic trajectory stays close to the original trajectory, suggesting its stability (gray). (b) The dynamics of entanglement entropy across different cuts on the periodic trajectory shows clear oscillations.}
\label{fig:TDVP}%
\vspace{-4mm}
\end{figure}

Such EOMs were previously obtained in the PXP model for $K=2$~\cite{Wenwei2018TDVP} and $K=3$~\cite{Michailidis2020Mixed} unit cells, and we obtain them for $K=4$ (see SM~\cite{SM}).
We focus on the $\ket{K}$ state from Eq.~(\ref{Eq:K-def}) with $\alpha=1$, $K=4$ parameterized by angles $\theta_j=(0,0,\pi/4,\pi/2)$ and its trajectory in the classical phase space. We find that this point is located exactly on a periodic trajectory, as shown by the solid red line in Fig.~\ref{fig:TDVP}(a). In addition, when perturbing the initial angles slightly, the resulting trajectory stays close to the periodic one at all times, indicating regular dynamics. A stability analysis of the periodic orbit confirms that it is stable with Lyapunov exponents being strictly zero up to numerical precision. Finally, the classical dynamics also allows us to compute various properties of the physical system, such as entanglement entropy, which is shown in Fig.~\ref{fig:TDVP}(b) for all nonequivalent bipartite cuts. The maximum entanglement entropy is $\ln(2)$, which occurs when the cut goes through the $W$ part of a $\ket{K}$ state.
This is in contrast with the PXP $\Zd{2}$ case, where the entanglement maximum occurs at the $\GSy$ state.

\begin{figure}[tb]
\centering
\includegraphics[width=\columnwidth]{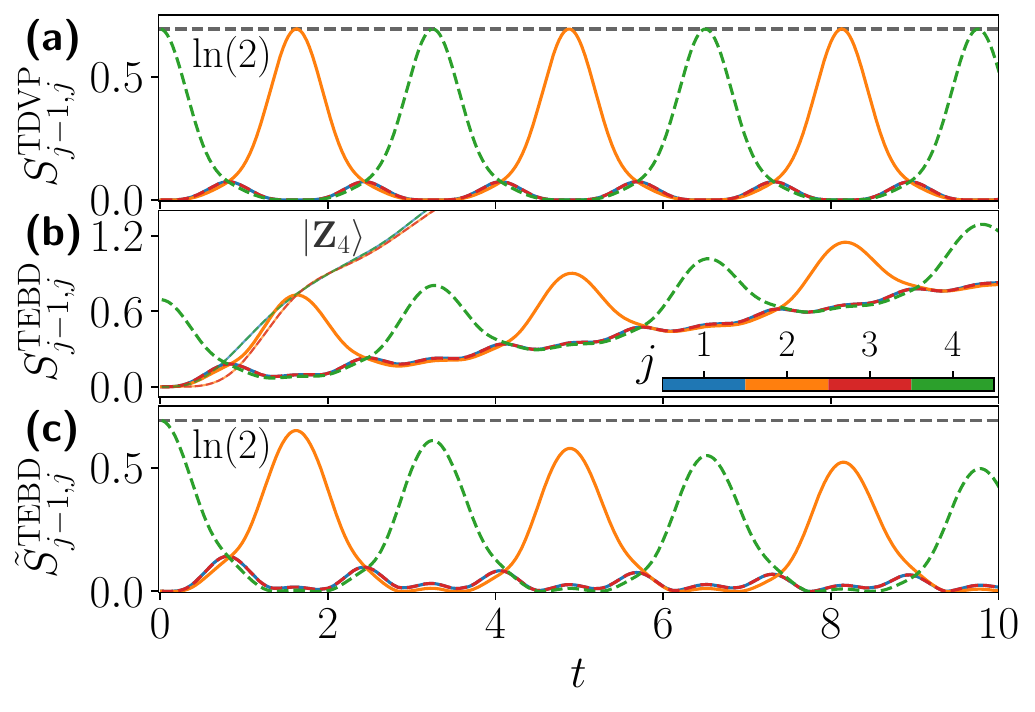}
\vspace{-7mm}
\caption{Entanglement dynamics after a quench from $\ket{K}$ with $\alpha=2$ and $K=4$. (a) The periodic trajectory in the projected dynamics predicts infinite oscillations of the entanglement entropy. (b) Exact dynamics obtained using TEBD~\cite{Vidal07,Schollwock} for $N=120$ sites, showing a linear ramp with oscillations on top. The growth of entanglement entropy is highly suppressed compared to the $\ket{\Zd{4}}$ case. (c) The entanglement entropy oscillations in the exact dynamics after subtracting 
 the linear ramp show good agreement with the results in (a).}
\label{fig:ent}%
\vspace{-4mm}
\end{figure}

Beyond the PXP model, we also find EOMs for the $\alpha=2$ and $K=4$ case (see SM~\cite{SM}), which also feature a periodic trajectory passing through the same initial point, $\theta_j=(0,0,\pi/4,\pi/2)$. Figure~\ref{fig:ent}(a) shows the entanglement dynamics on this trajectory. The exact entanglement dynamics in a long chain obtained with time-evolving block decimation (TEBD) method shown in Fig~\ref{fig:ent}(b) show oscillations with the same period but with an added slow linear growth. This highlights that the projection onto the variational manifold is not exact, as it only captures the periodic behavior but not its gradual decay. After subtracting the linear growth, the entanglement oscillations in Fig~\ref{fig:ent}(c) agree well with entanglement on the variational trajectory, Fig~\ref{fig:ent}(a). As for the PXP $K=4$ case, the maximum entanglement also occurs when a $\ket{K}$ state is cut in its $W$ part. We expect this behavior to be identical for all $\alpha$ as long as $K>2$.

\end{appendix}

\clearpage
\onecolumngrid
\begin{center}
\textbf{\large Supplemental Online Material for ``Quantum Many-Body Scars beyond the PXP model in Rydberg simulators" }\\[5pt]
Aron Kerschbaumer${}^1$, Marko Ljubotina${}^{1,2,3}$, Maksym Serbyn${}^1$, and Jean-Yves Desaules${}^1$ \\
{\small \sl ${}^1$Institute of Science and Technology Austria (ISTA), Am Campus 1, 3400 Klosterneuburg, Austria}\\
{\small \sl ${}^2$Physics Department, Technical University of Munich, TUM School of Natural Sciences,\\  Lichtenbergstr. 4,
Garching 85748, Germany}\\
{\small \sl ${}^3$Munich Center for Quantum Science and Technology (MCQST), Schellingstr. 4, München 80799, Germany}

\vspace{0.1cm}
\begin{quote}
{\small In this Supplementary Material, we provide additional information and data to support the results in the main text. In Sec.~\ref{SM:k}, we discuss the different symmetry sectors in which QMBSs appear and the degeneracies between scarred eigenstates.
In Sec.~\ref{SM:algebra}, we discuss the approximate $\mathfrak{su}(2)$ algebra for a few specific combinations of $\alpha$ and $K$, including cases with $K<2\alpha$ and $K>2\alpha+2$. In Sec.~\ref{SM:init}, we compare revivals from different initial states in the same system. In Sec.~\ref{SM:fid_dens}, we discuss the fidelity density ratio used in Fig.~3(b) of the main text and provide the raw fidelity data used to compute it as well as additional details. In Sec.~\ref{SM:scaling}, we discuss how our findings fare as the system size is increased. In Sec.~\ref{SM:qdim}, we show how to analytically compute the Hilbert space dimension for arbitrary $\alpha$, $N$ and boundary conditions. We use these results to compute the asymptotic quantum dimension. In Sec.~\ref{SM:data} we provide additional data for fidelity and entanglement entropy to complement the figures in the main text. Finally, in Sec.~\ref{SM:eom} we provide equations of motion for the variational angles in the semiclassical limit.}\\[10pt]
\end{quote}
\end{center}
\setcounter{equation}{0}
\setcounter{figure}{0}
\setcounter{table}{0}
\setcounter{page}{1}
\setcounter{section}{0}
\makeatletter
\renewcommand{\theequation}{S\arabic{equation}}
\renewcommand{\thefigure}{S\arabic{figure}}
\renewcommand{\thesection}{S\arabic{section}}
\renewcommand{\thepage}{\arabic{page}}
\renewcommand{\thetable}{S\arabic{table}}

\vspace{0cm}

\section{Symmetries and QMBS degeneracies}\label{SM:k}
While in the main text we have focused mostly on the dynamics, in this section we discuss some specific properties of the scarred eigenstates. These non-thermal eigenstates occur at $2N/K+1$ energies throughout the spectrum, however, the number of scarred eigenstates is generally not the same. We note that we only focus on the eigenstates that have the most prominent overlap with $\ket{K}$ among eigenstates with similar energies. We leave the study of states with smaller (but still large) overlaps with that state for future works. 

In order to count the number of scarred states, we must first discuss the symmetries of the Hamiltonian in Eq.~(1) of the main text. This operator is invariant under translation $T$ and under spatial inversion $\sr$ for all values of $\alpha$.
It follows that we can decompose the Hilbert space into disconnected sectors with fixed momenta, such that for any state $\ket{\psi}$ in the sector we have $T\ket{\psi}=e^{ik}\ket{\psi}$. As we are on a finite lattice where $T^{N}=\id$, we must have
$\ket{\psi}=T^N\ket{\psi}=e^{ikN}\ket{\psi}$. This implies that $e^{ikN}=1$ and therefore $k=n 2\pi/N$ with $n$ being an integer. Finally, as the lattice is discrete we can only translate by an integer number of sites, meaning that $k$ and $k+2\pi$ are indistinguishable. We thus restrict to $k=\in(-\pi,\pi]$, which translates into $n=-N/2+1$ to $N/2$ for $N$ even, and $n=-(N-1)/2$ to $(N-1)/2$ for $N$ odd.

The spatial inversion operator $\sr$ maps $k$ to $-k$. As such $T$ and $\sr$ do not generally commute and cannot be diagonalized simultaneously. Nonetheless, as $H$ commutes with both it means that sectors with momenta $k$ and $-k$ have identical energies and that their eigenstates are related by the action of $\sr$. Additionally, since our Hamiltonian is real, the eigenstates with momenta $+k$ and $-k$ are simply related by complex conjugation.
We also note the two special case $k=0$ and $k=\pi$ for which $k$ and $-k$ are equivalent. In these cases, $T$ and $\sr$ commute and we can divide these sectors into two subsectors with eigenvalue $p=\pm 1$ under $\sr$. 

The most symmetric sector is then the one with $k=0$ and $p=1$. Naturally, this is the sector in which the ground  state is found, which is also the first scarred state. Going up in energy, we find that the scarred eigenstates alternate between this sector and the ones with $k=\pm 2\pi/K$. This is not too surprising, as for a state $\ket{\psi_K}$ with periodicity $K$ these are the minimum allowed momenta beyond $k=0$. Indeed, for such a state $\ket{\psi}=T^K\ket{\psi}=e^{ikK}\ket{\psi}$, and so $k=2\pi m/K$ with $m$ an integer.

For $K=2$, $2\pi/K$ reduces to $\pi$ and we get $k=0$ and $k=\pi$. This is consistent with previous works on the PXP model that found that the $N+1$ scarred eigenstates alternate between the sectors $\{k=0,p=+1\}$ and $\{k=\pi,p=-1\}$~\cite{TurnerPRB}.
We note that this fact has been used to understand scars in this model as a condensate of magnons with $\pi$-momentum~\cite{IadecolaMagnons}.

\begin{figure}[ht]
\centering
\includegraphics[width=\columnwidth]{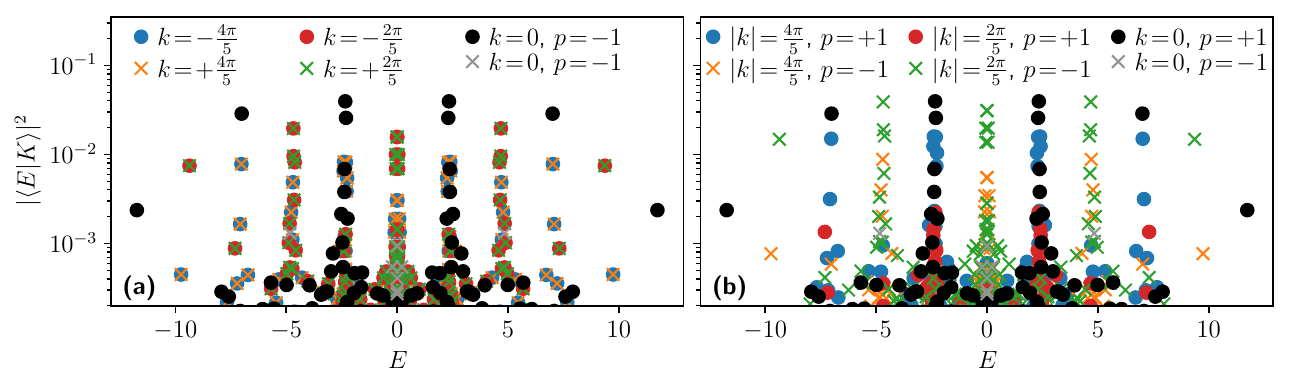}
\caption{Overlap of the $\ket{K}$ state with the eigenstates in all relevant symmetry sectors. The parameters are $\alpha=2$, $K=5$ and $N=25$. (a) Momentum is maximally resolved, showing the equal overlap of $\ket{K}$ with eigenstates with $\pm k$. (b) Spatial inversion is maximally resolved, breaking the degeneracy in the overlap with the $\ket{K}$ state. The ``towers of states'' also show clear alternation between $p=\pm 1$.
}
\label{fig:olap_kp}
\end{figure}

Interestingly, going beyond this simplest case of $K=2$ leads to a non-trivial difference. As soon as $K>2$, $k=2\pi /K$ and $k=-2\pi /K$ are physically different sectors. As their energies are identical and their eigenstates are related by the application of $\sr$, if one sector has QMBSs then the other sector must have scars as well. As such, we find that all QMBSs with $k\neq 0$ are doubly degenerate. Since the ground state has $k=0$ and the $2N/K+1$ scarred energies alternate between this value and $k=\pm 2\pi/K$, we can deduce that there are always $N/K+1$ scars with $k=0$, $N/K$ with $k=2\pi/N$ and $N/K$ with $k=-2\pi/N$ (assuming $N/K$ integer). So for $K>2$ we end up with a total of $3N/K+1$ eigenstates for $2N/K+1$ energies. Additionally, the $\ket{K}$ state must have equal overlap on the two degenerate scarred states. This simply follows from the fact that the eigenstates with $\pm k$ are related by complex conjugation. This is illustrated in Fig.~\ref{fig:olap_kp}(a). 

In order to break the degeneracy between the $k$ and $-k$ scarred eigenstates, one can try to instead use the so-called ``semi-momentum'' states for which $\sr$ is fully resolved but momentum is only resolved up to an absolute value. Essentially, this means that instead of separating the sectors with $|k|$ into $\pm k$, we separate them into $p=\pm 1$. The results of this procedure are shown in Fig.~\ref{fig:olap_kp}(b) for $\alpha=2$ and $K=5$. We now have a unique scarred eigenstate at the top of each ``tower of states''. Interestingly, these eigenstates (and the full towers) alternate between $p=+1$ and $p=-1$. This suggests that this semi-momentum picture might be more relevant when looking at QMBSs for $K>2$. It is also more intuitive with respect to the dynamics, as we focus on state transfer between two states which are related by the application of $\sr$, but not necessarily by translation (e.g. for $K$ odd).

\section{Algebra for specific $\alpha$ and $K$}\label{SM:algebra}
In this section, we discuss the approximate $\mathfrak{su}(2)$ algebra for a few specific cases of $\alpha$ and $K$ and give additional information such as the commutation relation between the various operators of the algebra. 

\subsection{$\alpha=1$ and $K=2$: a review of the algebra for the PXP $\Zd{2}$ scars}
We will focus on the definition introduced in Refs.~\cite{Choi2018su2,Bull2020Lie} which stems from the forward-scattering approximation (FSA)~\cite{TurnerNature,TurnerPRB}. This latter approach is based on separating the PXP Hamiltonian into a raising operator $\Jp$ and a lowering operator $\Jm$ as
\begin{equation}
    \Jp=\sum_{j=1}^{N/2} \PR_{2j-2}\spN_{2j-1}\PR_{2j}+\PR_{2j-1}\smN_{2j}\PR_{2j+1} \quad \text{and} \quad \Jm=\left(\Jp\right)^\dagger.
\end{equation}
An approximate basis for the scarred subspace can then be obtained as $\ket{n}=\frac{1}{\mathcal{N}_n}\left(\Jp\right)^n\ket{\Zd{2}}$, where $\mathcal{N}_n$ is a normalization factor. It is straightforward to check that $\ket{n=N}$ is simply the other $\Zd{2}$ state (i.e $\ket{\ua\da\ua\da\cdots \ua\da}$ and that applying $\Jp$ to it annihilates it. We thus end up with a set of $N+1$ basis states which capture the scarred dynamics and eigenstates despite composing a vanishing fraction of the Hilbert space.

Beyond approximating the dynamics, the decomposition of the Hamiltonian into $\Jp$ and $\Jm$ also allows us to define an $\mathfrak{su}(2)$ algebra from which we can get both the $\Jx$ and $\Jy$ operators, which can then be used to obtain $\Jz$. We end up with the following global $\mathfrak{su}(2)$ operators
\begin{equation}\label{eq:PXP_su2}
\begin{aligned}
    \Jx&=\frac{\Jp+\Jm}{2}=\frac{1}{2}\Ham_\mathrm{PXP}=\frac{1}{2} \sum_{j=1}^N \PR_{j-1}\sxN_j \PR_{j+1},\\
    \Jy&=\frac{\Jp-\Jm}{2i}=\, -\frac{1}{2} \sum_{j=1}^N (-1)^j\PR_{j-1}\syN_j \PR_{j+1}, \\
        \Jz&= \frac{1}{2}\left[\Jp,\Jm\right] = \frac{1}{i}[\Jx,\Jy] = -\frac{1}{2}\sum_{j=1}^N (-1)^j \PR_{j-1}\szN_j\PR_{j+1},
\end{aligned}
\end{equation}
and commutation relations
\begin{equation}\label{eq:PXP_su2_comms}
\begin{aligned}
    [\Jx,\Jy]&=i\Jz,\\
    [\Jy, \Jx]&=i\Jy{+}\frac{i}{4}\sum_{j=1}^N (-1)^j \left(\PR_{j-2}+\PR_{j+2}\right)\PR_{j-1}\syN_j \PR_{j+1},\\
    [\Jy,\Jz]&=i\Jx{+}\frac{i}{4}\sum_{j=1}^N \left(\PR_{j-2}+\PR_{j+2}\right)\PR_{j-1}\sxN_j \PR_{j+1}.
\end{aligned}
\end{equation}
As indicated by the presence of additional terms in these relations, the algebra is not exactly closed. However, we can still understand the scarred dynamics as the precession of a collective spin.
The two N\'eel states are respectively the ground and ceiling states of $\Jz$.
As the  Hamiltonian is directly proportional to $\Jx$, we will have a precession around the $x$ axis. Therefore the collective spin will rotate in the $yz$-plane and will explore the ground and ceiling states of both the $\Jz$ and $\Jy$ operators.

\subsection{$\alpha=1$ and $K=3$}
For the case with $\alpha=1$ and $K=3$, the raising operator is
\begin{equation}
    \Jp=\sum_j \PR_{3j}\spN_{3j+1}\PR_{3j+2}+\frac{1}{2}\PR_{3j+1}\sxN_{3j+2}\PR_{3j+3}+\PR_{3j+2}\smN_{3j+3}\PR_{3j+4},
\end{equation}
leading to the $\Jy=(\Jp-\Jm)/(2i)$ operator
\begin{equation}
    \Jy=\frac{1}{2}\sum_j \PR_{3j}\syN_{3j+1}\PR_{3j+2}-\PR_{3j+2}\syN_{3j+3}\PR_{3j+4}.
\end{equation}
This allows us to compute $\Jz=\frac{1}{i}[\Jx,\Jy]$, leading to
\begin{equation}
\begin{aligned}
    \Jz&=\frac{1}{2}\sum_j \PR_{3j}\szN_{3j+1}\PR_{3j+2}+\frac{1}{2}\PR_{3j}\left(\spN_{3j+1}\smN_{3j+2}+
    \smN_{3j+1}\spN_{3j+2}\right)\PR_{3j+3}\\
    &-\PR_{3j+2}\szN_{3j+3}\PR_{3j+4}-\frac{1}{2}\PR_{3j+1}\left(\spN_{3j+2}\smN_{3j+3}+
    \smN_{3j+2}\spN_{3j+3}\right)\PR_{3j+4}.
\end{aligned}
\end{equation}
To check how good the algebra is, we can compute $\frac{1}{i}[\Jz,\Jx]$ and see how close to $\Jy$ it is.
We end up with 
\begin{equation}
\begin{aligned}
    \frac{1}{i}[\Jz,\Jx]{=}\Jy{+}\frac{1}{8}\sum_j\bigg[& i\PR_{3j-1}(\spN_{3j}\smN_{3j+1}\spN_{3j+2}{-}\smN_{3j}\spN_{3j+1}\smN_{3j+2})\PR_{3j+3}{-}i\PR_{3j+1}(\spN_{3j+2}\smN_{3j+3}\spN_{3j+4}{-}\smN_{3j+2}\spN_{3j+3}\smN_{3j+4})\PR_{3j+5}\\
    &+3\PR_{3j}\PR_{3j+1}\syN_{3j+2}\PR_{3j+3}-
    3\PR_{3j+1}\syN_{3j+2}\PR_{3j+3}\PR_{3j+4}
    +\PR_{3j}\syN_{3j+1}\PR_{3j+2}\PR_{3j+3}\\
    &-2\PR_{3j-1}\PR_{3j}\syN_{3j+1}\PR_{3j+2}
    +2\PR_{3j+2}\syN_{3j+3}\PR_{3j+4}\PR_{3j+5}-\PR_{3j+1}\PR_{3j+2}\syN_{3j+3}\PR_{3j+4}\bigg]
\end{aligned}
\end{equation}
While we do recover $\Jy$, there are additional terms as for $K=2$ in the same model.

\subsection{$\alpha=1$ and $K=4$}
We now look at the $K=4$ case for the PXP model. We start by defining
\begin{equation}
\begin{aligned}
        \Jp=\sum_j \PR_{4j}\spN_{4j+1}\PR_{4j+2}+\PR_{4j+1}\spN_{4j+2}\PR_{4j+3}+\PR_{4j+2}\smN_{4j+3}\PR_{4j+4}+ \PR_{4j+3}\smN_{4j+4}\PR_{4j+5},
\end{aligned}
\end{equation}
which leads to
\begin{equation}
    \Jy=\frac{1}{2}\sum_j \PR_{4j}\syN_{4j+1}\PR_{4j+2}+\PR_{4j+1}\syN_{4j+2}\PR_{4j+3}-\PR_{4j+2}\syN_{4j+3}\PR_{4j+4}- \PR_{4j+3}\syN_{4j+4}\PR_{4j+5}
\end{equation}
and
\begin{equation}
\begin{aligned}
        \Jz&=\frac{1}{2}\sum_j \bigg[\PR_{4j}\szN_{4j+1}\PR_{4j+2}+\PR_{4j+1}\szN_{4j+2}\PR_{4j+3}-\PR_{4j+2}\szN_{4j+3}\PR_{4j+4}- \PR_{4j+3}\szN_{4j+4}\PR_{4j+5}\\
        &+\PR_{4j}\left(\spN_{4j+1}\smN_{4j+2}+\smN_{4j+1}\spN_{4j+2}\right)\PR_{4j+3}-\PR_{4j+2}\left(\spN_{4j+3}\smN_{4j+4}+\smN_{4j+3}\spN_{4j+4}\right)\PR_{4j+5}\bigg].
\end{aligned}
\end{equation}
We note that the ground state of this operator is \emph{exactly} the $\ket{K}$ state, thus providing a self-consistency check of our algebraic approach.
Finally, we check the closure of the algebra by computing 
\begin{equation}
\begin{aligned}
    \frac{1}{i}[\Jz,\Jx]=\Jy{+}\frac{1}{4}\sum_j& \bigg[i\PR_{4j-1}(\spN_{4j}\smN_{4j+1}\spN_{4j+2}-\smN_{4j}\spN_{4j+1}\smN_{4j+2})\PR_{4j+3}-\PR_{4j-1}\syN_{4j}\PR_{4j+1}\PR_{4j+2}\\
    &-i\PR_{4j}(\spN_{4j+1}\smN_{4j+2}\spN_{4j+3}-\smN_{4j+1}\spN_{4j+2}\smN_{4j+3})\PR_{4j+4}-\PR_{4j+1}\PR_{4j+2}\syN_{4j+3}\PR_{4j+4}\\
    &+i\PR_{4j+1}(\spN_{4j+2}\smN_{4j+3}\spN_{4j+4}-\smN_{4j+2}\spN_{4j+3}\smN_{4j+4})\PR_{4j+5}+\PR_{4j+1}\syN_{4j+2}\PR_{4j+3}\PR_{4j+4}\\
    &+i\PR_{4j+2}(\spN_{4j+3}\smN_{4j+4}\spN_{4j+5}-\smN_{4j+3}\spN_{4j+4}\smN_{4j+5})\PR_{4j+6}+\PR_{4j-1}\PR_{4j}\syN_{4j+1}\PR_{4j+2}\\
    &-2\PR_{4j}\PR_{4j+1}\syN_{4j+2}\PR_{4j+3}-2\PR_{4j}\syN_{4j+1}\PR_{4j+2}\PR_{4j+3}+2\PR_{4j+2}\PR_{4j+4}\syN_{4j+4}\PR_{4j+5}\\
    &+2\PR_{4j+2}\syN_{4j+3}\PR_{4j+4}\PR_{4j+5}\bigg].
\end{aligned}
\end{equation}

\subsection{$K>2\alpha+2$: $\alpha=1$ and $K=5$}
In order to illustrate what happens to the algebra when $K>2\alpha+2$, we will look at the case with $\alpha=1$ and $K=5$.
We start with the $\Jp$ operator defined as 
\begin{equation}
\begin{aligned}
        \Jp=\sum_j \PR_{5j}\spN_{5j+1}\PR_{5j+2}+\PR_{5j+1}\spN_{5j+2}\PR_{5j+3}+\frac{1}{2}\PR_{5j+2}\sxN_{5j+3}\PR_{5j+4}+\PR_{5j+3}\smN_{5j+4}\PR_{5j+5}+ \PR_{5j+4}\smN_{5j+5}\PR_{5j+6},
\end{aligned}
\end{equation}
from which we get
\begin{equation}
    \Jy=\frac{1}{2}\sum_j \PR_{5j}\syN_{5j+1}\PR_{5j+2}+\PR_{5j+1}\syN_{5j+2}\PR_{5j+3}-\PR_{5j+3}\syN_{5j+4}\PR_{5j+5}-\PR_{5j+4}\syN_{5j+5}\PR_{5j+6}.
\end{equation}
and
\begin{equation}
\begin{aligned}
    \Jz=\frac{1}{2}\sum_j &\PR_{5j}\szN_{5j+1}\PR_{5j+2}+\PR_{5j+1}\szN_{5j+2}\PR_{5j+3}-\PR_{5j+3}\szN_{5j+4}\PR_{5j+5}-\PR_{5j+4}\szN_{5j+5}\PR_{5j+6}\\
    &+\PR_{5j}\left(\spN_{5j+1}\smN_{5j+2}+\smN_{5j+1}\spN_{5j+2}\right)\PR_{5j+3}+\frac{1}{2}\PR_{5j+1}\left(\spN_{5j+2}\smN_{5j+3}+\smN_{5j+2}\spN_{5j+3}\right)\PR_{5j+4}\\
    &-\frac{1}{2}\PR_{5j+2}\left(\spN_{5j+3}\smN_{5j+4}+\smN_{5j+3}\spN_{5j+4}\right)\PR_{5j+5}-\PR_{5j+3}\left(\spN_{5j+4}\smN_{5j+5}+\smN_{5j+4}\spN_{5j+5}\right)\PR_{5j+6}.
\end{aligned}
\end{equation}
Unlike in the case with $2\alpha \leq K \leq 2\alpha+2$, there is no $XY$-type term between sites $5j+1$ and $5j+3$ (and between $5j+3$ and $5j+5$). This is generically what happens for $K>2\alpha+2$ and it leads to changes in the ground state of $\Jz$. Essentially, in the second half of each unit-cell instead of a $\ket{W}$ state the superposition is now unequal, with the middle sites having a much higher probability of being occupied as they are involved in more $XY$-type terms. This breaks the self-consistency of our approach, since this unequal structure was not present in our Ansatz $\Jp$.

Additionally, it is clear that applying $\Ham_\alpha$ to this unequal superposition (or to $\ket{K}$) will not only take it to the $\ket{0}$ state but can also create additional excitations. This leads to an overlap on states like $\ket{\da\da\ua\da\ua}$ for the case $\alpha=1$ and $K=5$. This completely kills the approximation that each half-cell is acting as a two-level system and it is not too surprising that the revivals are destroyed.

We can also look at closure of the algebra as
\begin{equation}
\begin{aligned}
    \frac{1}{i}[\Jz,\Jx]=\Jy{+}\frac{1}{4}\sum_j& i\PR_{5j-1}(\spN_{5j}\smN_{5j+1}\spN_{5j+2}-\smN_{5j}\spN_{5j+1}\smN_{5j+2})\PR_{5j+3}-\PR_{5j-1}\syN_{5j}\PR_{5j+1}\PR_{5j+2}\\
    -&i\frac{3}{2}\PR_{5j}(\spN_{5j+1}\smN_{5j+2}\spN_{5j+3}-\smN_{5j+1}\spN_{5j+2}\smN_{5j+3})\PR_{5j+4}-2\PR_{5j}\PR_{5j+1}\syN_{5j+2}\PR_{5j+3}\\
    +&i\frac{3}{2}\PR_{5j+2}(\spN_{5j+3}\smN_{5j+4}\spN_{5j+5}-\smN_{5j+3}\spN_{5j+4}\smN_{5j+5})\PR_{5j+6}-2\PR_{5j}\syN_{5j+1}\PR_{5j+2}\PR_{5j+3}\\
    +&i\PR_{5j+3}(\spN_{5j+4}\smN_{5j+5}\spN_{5j+6}-\smN_{5j+5}\spN_{5j+6}\smN_{5j+6})\PR_{5j+7}+\frac{3}{2}\PR_{5j+1}\PR_{5j+2}\syN_{5j+3}\PR_{5j+4}\\
    -&\frac{1}{2}\PR_{5j+1}\syN_{5j+2}\PR_{5j+3}\PR_{5j+4}+\frac{3}{2}\PR_{5j+2}\syN_{5j+3}\PR_{5j+4}\PR_{5j+5}+\frac{1}{2}\PR_{5j+2}\PR_{5j+3}\syN_{5j+4}\PR_{5j+5}\\
    +&2\PR_{5j+3}\syN_{5j+4}\PR_{5j+5}\PR_{5j+6}+2\PR_{5j+3}\PR_{5j+4}\syN_{5j+5}\PR_{5j+6}+\PR_{5j-1}\PR_{5j}\syN_{5j+1}\PR_{5j+2}.
\end{aligned}
\end{equation}
While there are more terms than in the $K=4$ case, they are of the same form and with similar prefactors. So from this point of view it is difficult to assess the quality of the algebra. Indeed, only the projection of these additional terms in the scarred subspace is relevant. However, this quantity is much harder to compute as it requires a way to construct (or at least approximate) this subspace.

\subsection{$K<2\alpha$: $\alpha=2$ and $K=3$}
To investigate what happens when $K<2\alpha$, we study the case with $\alpha=2$ and $K=3$. We start from the definition of the raising operator as
\begin{equation}
    \Jp=\sum_j \PR_{3j-1}\PR_{3j}\spN_{3j+1}\PR_{3j+2}\PR_{3j+3}+\frac{1}{2}\PR_{3j}\PR_{3j+1}\sxN_{3j+2}\PR_{3j+3}\PR_{3j+4}+\PR_{3j+1}\PR_{3j+2}\smN_{3j+3}\PR_{3j+4}\PR_{3j+5},
\end{equation}
from which we compute
\begin{equation}
    \Jy=\frac{1}{2}\sum_j \PR_{3j-1}\PR_{3j}\syN_{3j+1}\PR_{3j+2}\PR_{3j+3}-\PR_{3j+1}\PR_{3j+2}\syN_{3j+3}\PR_{3j+4}\PR_{3j+5}
\end{equation}
and
\begin{equation}
\begin{aligned}
        \Jz=&\frac{1}{2}\sum_j \PR_{3j-1}\PR_{3j}\szN_{3j+1}\PR_{3j+2}\PR_{3j+3}-\PR_{3j+1}\PR_{3j+2}\szN_{3j+3}\PR_{3j+4}\PR_{3j+5}\\
        &+\frac{1}{4}\sum_j\bigg[ \PR_{3j}\PR_{3j+1}\left(\spN_{3j+2}\PR_{3j+3}\smN_{3j+4}+\smN_{3j+2}\PR_{3j+3}\spN_{3j+4}\right)\PR_{3j+5}\PR_{3j+6}\\
        &-\PR_{3j+1}\PR_{3j+2}\left(\spN_{3j+3}\PR_{3j+4}\smN_{3j+5}+\smN_{3j+4}\PR_{3j+4}\spN_{3j+5}\right)\PR_{3j+6}\PR_{3j+7}\\
        &+\PR_{3j-1}\PR_{3j}\left(\spN_{3j+1}\smN_{3j+2}+\smN_{3j+1}\spN_{3j+2}\right)\PR_{3j+3}\PR_{3j+4}-\PR_{3j}\PR_{3j+1}\left(\spN_{3j+2}\smN_{3j+3}+\smN_{3j+2}\spN_{3j+3}\right)\PR_{3j+4}\PR_{3j+5}\bigg].
\end{aligned}
\end{equation}
In this case, we see the appearance of terms of the form $\PR\PR(\spN\PR\smN+\smN\PR\spN)\PR\PR$ obtained from the commutation of $\PR\PR \sxN_j \PR\PR$ and $\PR\PR \syN_{j+2} \PR\PR$. These $XY$-type terms then connect different cells together. For $K=4$ similar terms would appear, but there would always be another contribution canceling it.

These unwanted $XY$ terms also appear for higher $\alpha$ as long as $K<2\alpha$. However, the ground state of $\Jz$ still locally looks like the $\ket{K}$ state. The main differences with the case $2\alpha \leq K\leq 2\alpha+2$ are twofold. First of all, the $\ket{W}$ part now has some different weights depending on where the excitation is. Additionally, the different unit cells now have a relatively large amount of entanglement. Indeed, placing an excitation at the last site of a unit cell blocks the first sites of the second half of the next cell. In the most extreme cases $K=\alpha+1$, the ground state is simply the $\Zd{\alpha+1}$ state. But for large enough $\alpha$ and $K$ close enough to $2\alpha$, the entanglement between cells becomes small and the algebraic structure is good enough to allow for revivals.

\section{Comparison between reviving initial states}\label{SM:init}
In this section, we discuss the difference in fidelity when quenching from different reviving states in the same system.
\subsection{General case}
\begin{figure}[htb]
\centering
\includegraphics[width=\columnwidth]{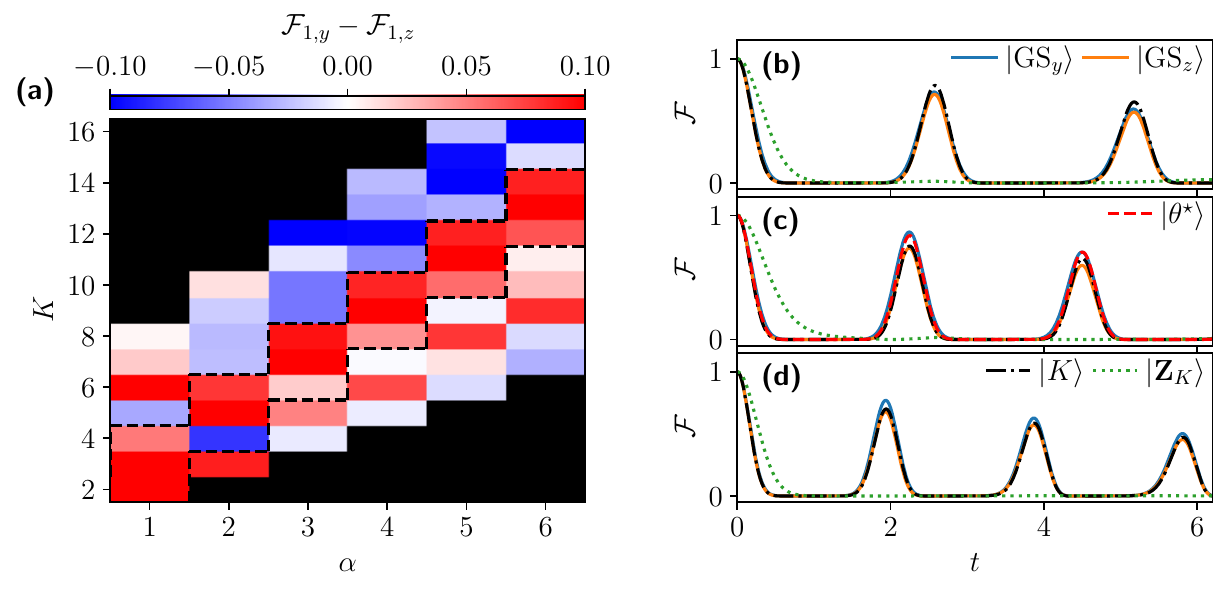}
\caption{(a) Difference in fidelity at the first revival between the $\GSy$ and $\GSz$ states. The $\GSy$ state shows better revivals in the vast majority of cases. (b)-(d) Fidelity after quenches from different initial states for $\alpha=3$ with (b) $K=6$ and $N=30$ (c) $K=7$ and $N=28$, and (d) $K=8$ and $N=32$. The state $\ket{\theta^\star}$ is only defined for $K=7$ and is characterized by angles $\theta^\star/\pi=(0, 0, 0, 0.0645, 0.2017, 0.2665, 0.5)$ in the MPS manifold.
}
\label{fig:LR_YZ}%
\end{figure}

First, we compare the ground state $\ket{{\rm GS}_y}$ of $\Jy$ and the ground state $\ket{{\rm GS}_z}$ of $\Jz$.
We find that in essentially all cases, the first revival from $\ket{{\rm GS}_y}$ is better, as showcased in Fig.~\ref{fig:LR_YZ}(a). For a few specific cases we also plot the time dynamics in Fig.~\ref{fig:LR_YZ}(b)-(d). Overall, we find that the difference is the strongest for $K=2\alpha+1$.

Nonetheless, as the trajectory from $\GSy$ passes close to $\GSz$, we know that there should be a state very similar to $\GSz$ with revivals comparable to that of $\GSy$. In order to find a simple suitable candidate, we restrict our search to the $\chi=\alpha+1$ MPS manifold introduced in Appendix~\ref{app:TDVP} of the main text. We fix the periodicity to $K=2\alpha+1$ and, to force a similar structure, we set the first angle to 0 and the last one to $\pi/2$. We then optimize for revival fidelity. We denote the resulting state by $\ket{\theta^\star}$. For $\alpha=1$ to 3, we find that the angles defining $\ket{\theta^\star}$ are respectively 
\begin{equation}
\theta^\star_{\alpha=1}/\pi=(0, 0.1162, 0.5), \ 
\theta^\star_{\alpha=2}/\pi=(0, 0, 0.1266, 0.25, 0.5) \ \text{and} \ \theta^\star_{\alpha=3}/\pi=(0, 0, 0, 0.0645, 0.2017, 0.2665, 0.5). 
\end{equation}
For comparison, the $K$ state with $\beta=0.65$ defined in the main text leads to the angles 
\begin{equation}
    \theta_{\alpha=1}/\pi=(0, 0.1835, 0.5), \ 
\theta_{\alpha=2}/\pi=(0, 0, 0.1371, 0.25, 0.5) \ \text{and} \ \theta_{\alpha=3}/\pi=(0, 0, 0, 0.1143, 0.1959, 0.25, 0.5).
\end{equation}

\begin{figure}[htb]
\centering
\includegraphics[width=\columnwidth]{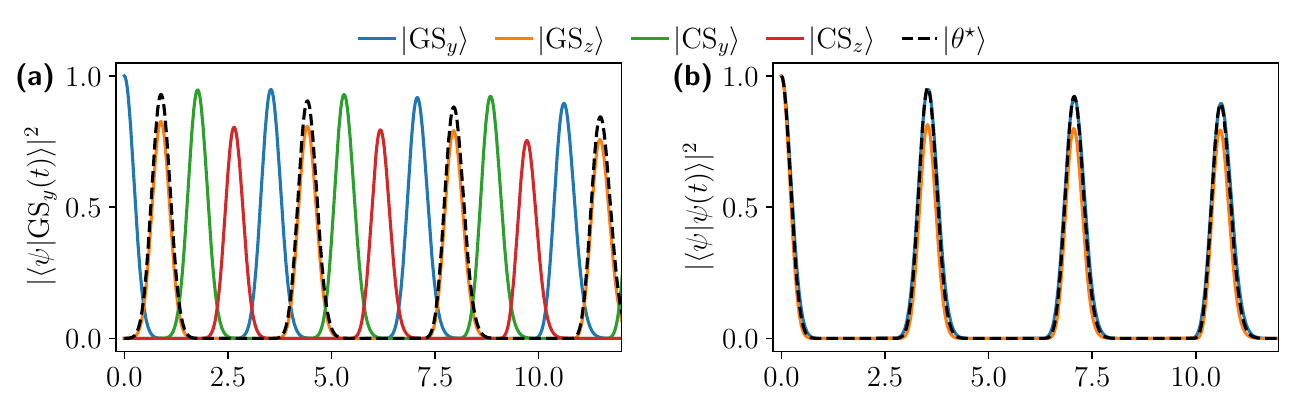}
\caption{Dynamics for $\alpha=1$, $K=3$ and $N=18$, with (a) state transfer after a quench from $\GSy$ and (b) revival fidelity after a quench from several states. The state $\ket{\theta^\star}$ was found by optimizing for first-revival fidelity over states in the MPS manifold and is parameterized by angles $\theta^\star/\pi=(0, 0.1162, 0.5)$. Its revivals are comparable to that of $\GSy$. From panel (a), it also lies closer to the trajectory of $\GSy$ than $\GSz$.
}
\label{fig:dyn_R1K3}%
\end{figure}

These optimized states then have revivals comparable to that of the $\GSy$ state, as shown in Fig.~\ref{fig:LR_YZ}(c) for $\alpha=3$ and $K=7$. We also show this for $\alpha=1$ and $K=3$ in Fig.~\ref{fig:dyn_R1K3}. In addition, on this figure we show that state transfer from the $\GSy$ state to $\ket{\theta^\star}$ is clearly higher than to $\GSz$, confirming that $\ket{\theta^\star}$ is on the right trajectory. This suggests that for $K=2\alpha+1$ the $\Jz$ operator of the algebra should be corrected in order to have $\ket{\theta^\star}$ as its ground state. We leave this task for future works.

\subsection{PXP model}
\begin{figure}[htb]
\centering
\includegraphics[width=\columnwidth]{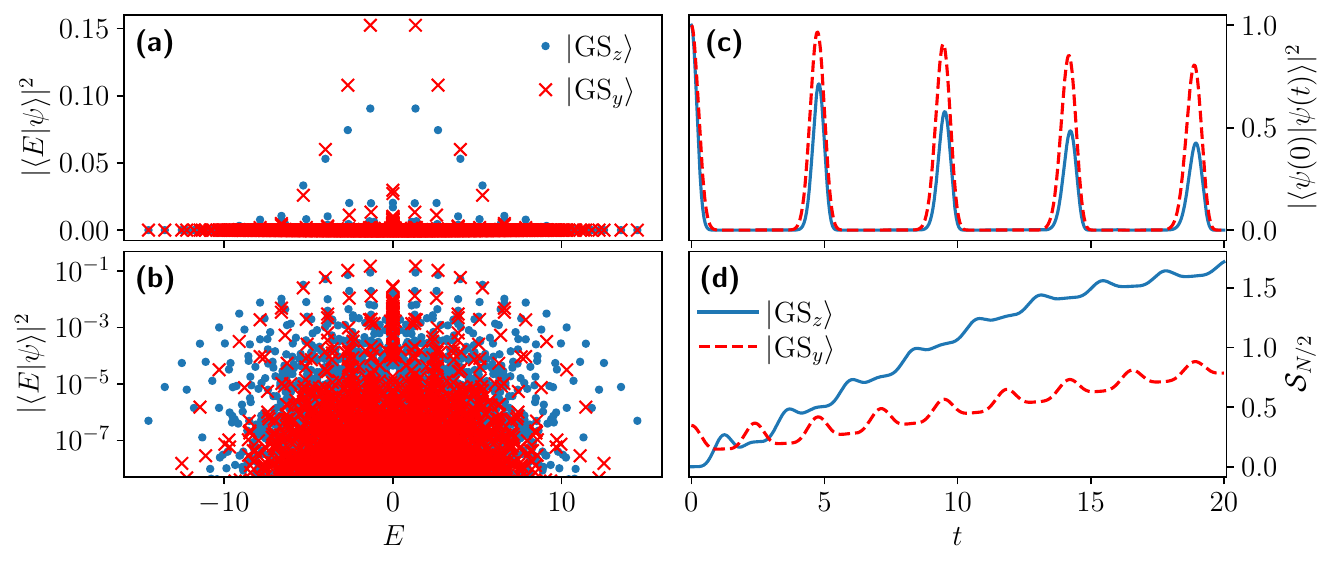}
\caption{Scarring for $\alpha=1$, $K=2$ and $N=24$. The $\GSz$ state is the N\'eel state that has been widely studied. Nonetheless, the $\ket{{\rm GS}_y}$ state demonstrates better revivals both in terms of fidelity and entropy growth. This is explained by a higher overlap with the scarred eigenstates. 
}
\label{fig:PXP_YZ}
\end{figure}

Interestingly, in the PXP case for $K=2$ we also find better revivals from $\GSy$ than from $\GSz=\ket{\Zd{2}}$, see Fig.~\ref{fig:PXP_YZ}. While both states have overlaps with the same scarred states, the former has a higher total overlap. For $N=24$, we find that the total overlap of $\GSz$ on scarred states and zero modes is 0.674. Meanwhile, for $\GSy$ this overlap is 0.897. On top of this, as clearly demonstrated in Fig.~\ref{fig:PXP_YZ}(a) and (b), $\GSy$ is more concentrated on the scarred states near $E=0$. As the spacing of these eigenstates gets smaller towards the edges of the spectrum, concentrating on states with small $|E|$ means that the energy spacing of the relevant states will be more uniform. This will slow down dephasing between them and thus make the dynamics periodic for a longer time.

Our approach also explains why $\ket{\Zd{3}}$ shows revival in the PXP model while $\ket{\Zd{4}}$ does not. In both cases, we can look at the overlap between the $\ket{\Zd{d}}$ state and the $\GSz$ state that revives due to the algebra. For $\ket{\Zd{3}}$, that state is $\ket{\theta^\star}$, which is parameterized by angles $(0,0.1162\pi,\pi/2)$ in the MPS manifold. As discussed above, this state revives better than the ground state of the theorized $\Jz$ operator, and shows larger state transfer to the ground state of $\Jy$. Its overlap with the $\ket{\Zd{3}}$ state is then simply $\left[\cos(0.1162\pi)\right]^{(2N/3)}\approx 0.9556^N$ for $N$ multiples of 3. For $\ket{\Zd{4}}$, the relevant state is $\ket{K=4}$, which is both the best reviving state in the manifold as well as the ground state of $\Jz$. Its overlap with $\ket{\Zd{4}}$ is  $\left(1/2\right)^{(N/4)}\approx 0.8409^N$ for $N$ multiples of 4. While these two numbers seem similar, for $N=12$ we get $0.5796$ for $\ket{\Zd{3}}$ but only $0.125$ for $\ket{\Zd{4}}$. As we reach $N=24$, the difference is even more striking, with an overlap of $0.3360$ for the former but $0.01563$ for the latter. 

\begin{figure}[hbt]
\centering
\includegraphics[width=\columnwidth]{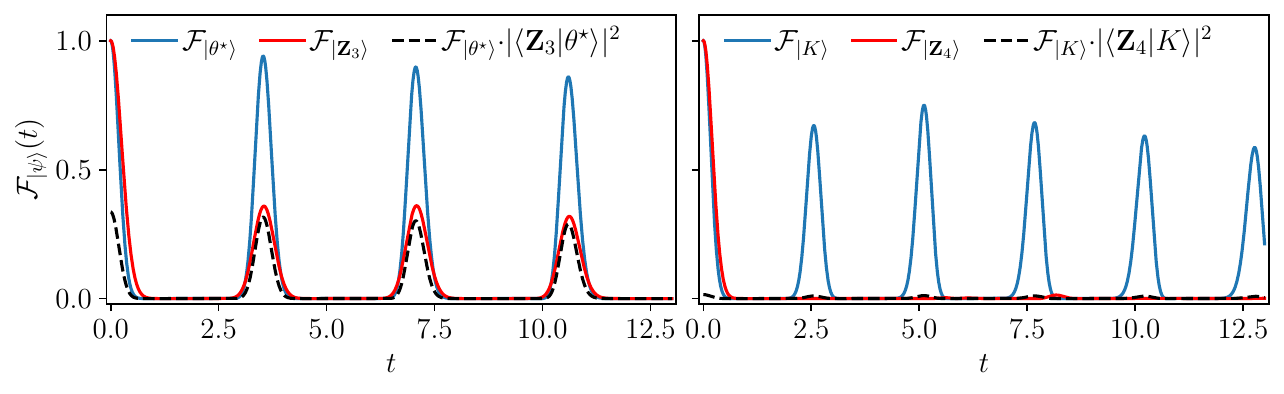}
\caption{Fidelity $\mathcal{F}$ after a quench from various states $\ket{\psi}$ in the PXP model ($\alpha=1$) for size-3 and -4 unit cells and $N=24$. In both cases, the approximation that the revivals of the $\ket{\Zd{d}}$ are explained by their overlap with the best reviving state provides a reasonable estimate of the actual revival amplitude.
}
\label{fig:revs_Z3_Z4}
\end{figure}

This allows us to devise a very crude approximation, in which we assume that all the oscillatory dynamics of the $\ket{\Zd{d}}$ state comes from its overlap with the $\ket{\theta^\star}$ or $\ket{K}$ state. This implies that the revivals of these states multiplied by their overlap with $\ket{\Zd{d}}$ should approximate the revivals of $\ket{\Zd{d}}$. We show results for this with $N=24$ in Fig.~\ref{fig:revs_Z3_Z4}. Despite how rough this approximation is, it provides a relatively good approximation of the revival amplitude of both $\ket{\Zd{3}}$ and $\ket{\Zd{4}}$. Therefore our results allow us to elucidate the mystery of why some $\ket{\Zd{d}}$ states revive in the PXP model and others do not, as all that is needed to determine this is their overlap with the extremal states of the algebra operators.
Based on these considerations, we also do not expect $\ket{\Zd{d}}$ type states to revive for higher $\alpha$. Indeed, their overlap with the relevant $\ket{K}$-type states gets smaller and smaller as $\alpha$ (and so the relevant values of K) increases. 

\section{Additional data for Fig.~3}\label{SM:fid_dens}
In Fig.~3(b) of the main text, we use $\ln(d)/f_1$ as a probe for ergodicity breaking across different $N$ and $\alpha$. In this section we provide additional details as to why this quantity was chosen and we show the raw fidelity data for comparison.
We remind that $f_1$ is the fidelity density at the first revival $f_1=-\ln(\mathcal{F}_1)/N$, where $\mathcal{F}_1$ is simply the fidelity at the first revival. Unlike $\mathcal{F}_1$, $f_1$ quickly converges to a finite value as $N$ is increased (see Sec.~\ref{SM:scaling}), and thus allows for the comparison of different system sizes.
However, this does not take into account how constrained the different models are. Indeed, after quenching from a random state we expect the wavefunction to spread evenly in the Hilbert space leading to a fidelity of $\approx 1/\mathcal{D}=1/d^N$, where $\mathcal{D}$ is the Hilbert space dimension and $d$ is the local quantum dimension that we compute in Sec.~\ref{SM:qdim}. This means that the corresponding fidelity density is $-\ln(1/d^N)/N=\ln(d)$. Hence $\ln(d)/f_1$ should be around 1 for a thermalizing state. As $f_1$ gets smaller the revivals improve (with perfect scarring leading to $f_1=0$), therefore a much larger value of $\ln(d)/f_1$ means anomalously high revivals and scarring. 

\begin{figure}
    \centering
\includegraphics[width=\columnwidth]{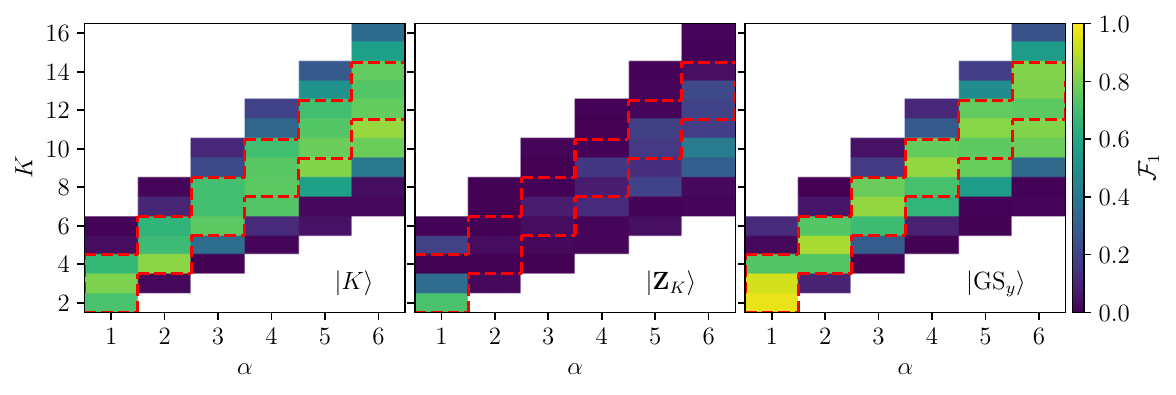}
\caption{Fidelity at the first revival for various initial states and blockade range}
\label{fig:fid_grid}
\end{figure}

To complement the rescaled fidelity density data in Fig.~3 of the main text, on Fig.~\ref{fig:fid_grid} we show the raw fidelity data for the same states and blockade ranges. The same patterns as in Fig.~3 of the main text can be seen.
We also provide the system size used for every combination of $\alpha$ and $K$ as well as the corresponding Hilbert space dimension (see Tab.~\ref{tab:ND_grid}).

\begin{table}
    \begin{tabular}{|l||c|c|c|c|c|c|}
 \hline
 16 & & &  &  &  & 48 \\
 \hline
 15 & & &  &  &  & 45 \\
 \hline
 14 & & &  &  & 42 & 42 \\
 \hline
 13 & & &  &  & 39 & 52 \\
 \hline
 12 & & &  & 36 & 48 & 48 \\
 \hline
 11 & & &  & 44 & 44 & 44 \\
 \hline
 10 & & & 30 & 40 & 40 & 50 \\
 \hline
 9 & & & 36 & 36 & 45 & 45 \\
  \hline
 8 & & 32 & 32 & 40 & 48 & 48 \\
  \hline
 7 & & 28 & 35 & 42 &49 & 49 \\
 \hline
 6 & 24 & 30 & 36 & 42 & 48 &  \\
 \hline
 5 & 25 & 30 & 35 & 40 &  &  \\
 \hline 
 4 & 24 & 32 & 36 &  &  &  \\
 \hline
 3 & 24 & 30 &  &  &  &  \\
 \hline
 2 & 24 &  &  &  &  &  \\
 \hhline{|=#=|=|=|=|=|=|}
 $K/\alpha$ & 1 & 2 & 3  & 4  & 5 & 6 \\
 \hline
\end{tabular}
\quad \quad \quad \quad
\begin{tabular}{|l||c|c|c|c|c|c|}
 \hline
 16 &  &  &  &  &  & 55181 \\
 \hline
 15 &  &  &  &  &  & 27895 \\
 \hline
 14 &  &  &  &  & 37737 & 14113 \\
 \hline
 13 &  &  &  &  & 17772 & 137138 \\
 \hline
 12 &  &  &  &  24916 & 170043 & 55181 \\
 \hline
 11 &  &  &  & 236281 &  62338 & 22232 \\
 \hline
 10 &  &  &  15812 &  76724  & 22839 & 87011 \\
 \hline
 9 &  &  & 109343 &  24916  & 80110 & 27895 \\
 \hline
 8 &  & 205221 & 30125 &  76724 & 170043 & 55181 \\
 \hline
 7 &  & 44483 & 79218 & 134645 & 218548 & 69294 \\
 \hline
 6 & 103682 & 95545 & 109343 & 134645 & 170043 &  \\
 \hline
 5 & 167761 & 95545 & 79218 & 76724 &  &  \\
 \hline
 4 & 103682 & 205221 & 109343 &  &  &  \\
 \hline
 3 & 103682 & 95545 &  &  &  &  \\
 \hline
 2 & 103682 &  &  &  &  &  \\
 \hhline{|=#=|=|=|=|=|=|}
  $K/\alpha$ & 1 & 2 & 3  & 4  & 5 & 6 \\
 \hline
    \end{tabular}
    \caption{System size (left) and Hilbert space dimension (right) for the data shown in Fig.~3(b) of the main text, in Fig.~\ref{fig:LR_YZ}(a) and in Fig.~\ref{fig:fid_grid}}
    \label{tab:ND_grid}
\end{table}

\section{System-size scaling}\label{SM:scaling}
\begin{figure}
    \centering
\includegraphics[width=\columnwidth]{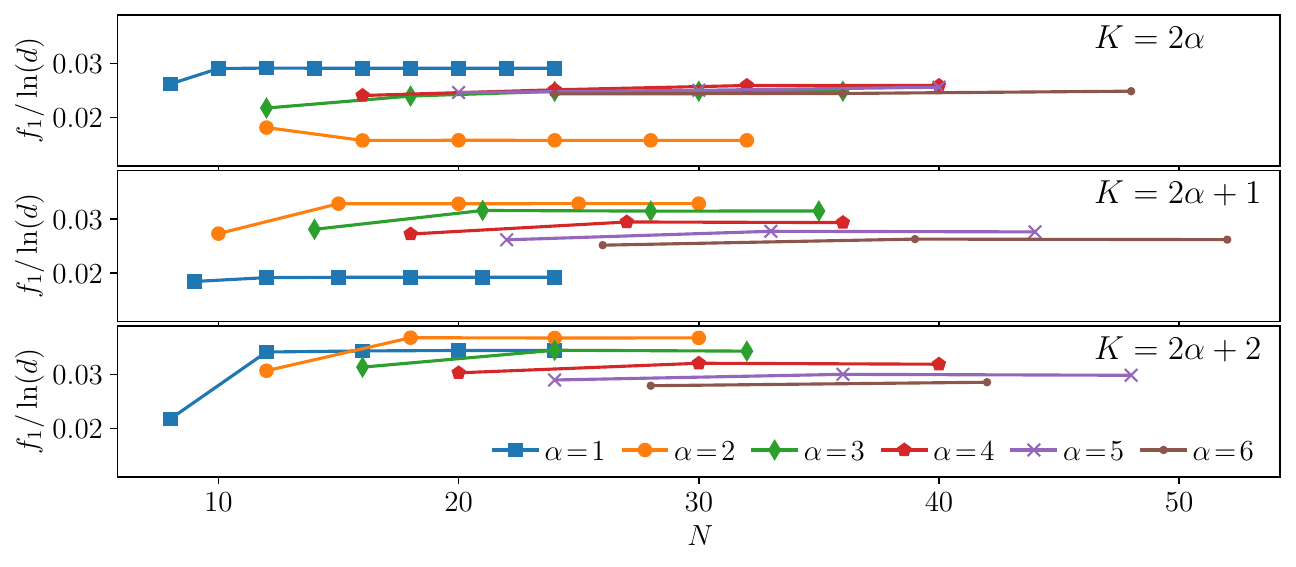}
\caption{Fidelity density at the first revival after a quench from the $\ket{K}$ state for various initial states, blockade ranges and system sites. The raw fidelity density $f_1$ is divided by $\ln(d)$ to account for the different level of constraints between systems. Overall, we see a good converge of $f_1$ with $N$.}
\label{fig:fid_scale}
\end{figure}

In this section, we briefly discuss the finite size scaling of dynamics and eigenstate properties with $N$. On Fig.~\ref{fig:fid_scale} we show the fidelity density at the first revival after a quench from the $K$ states with $K=2\alpha$ to $K=2\alpha+2$ for a range of values of $\alpha$ and $N$. The fidelity density is divided by $\ln(d)$, with $d$ the local quantum dimension (see Sec.~\ref{SM:qdim}), to take into account the fact that different $\alpha$ lead to different levels of constraints. Overall, for all $\alpha$ a clear convergence towards a finite value of $f_1$ is seen, as is the case for the PXP model ($\alpha=1$). Nonetheless, as $\alpha$ is increased, larger system sizes are required to reach the regime where finite-size effects are negligible.

As the rescaling of $f_1$ by $\ln(d)$ affects all data-points for a given $\alpha$ identically, it does not warp the convergence of $f_1$ in any way. However, it allows for a more meaningful comparison of the strength for scarring between various $\alpha$. In particular, for $K=2\alpha$ it appears that $f_1/\ln(d)$ is already converged as a function of  $\alpha$ for $\alpha=3$. Based on this data, we conjecture that for a large enough $N$ all $\alpha$ show the same value of $f_1/\ln(d)$. While the same is likely true for $K=2\alpha+1$ and $2\alpha+2$, the data collected is not conclusive. However, as a lower $f_1$ indicates better revivals, it means that scarring is \emph{stronger} as $\alpha$ is increased. The fact that not only $f_1$ but also $f_1/\ln(d)$ gets smaller with $\alpha$ shows that scarring is not getting stronger simply due to a decreasing Hilbert space size scaling, but that the effect is genuine and thus remains present after appropriate rescaling.

\begin{figure}
    \centering
\includegraphics[width=\columnwidth]{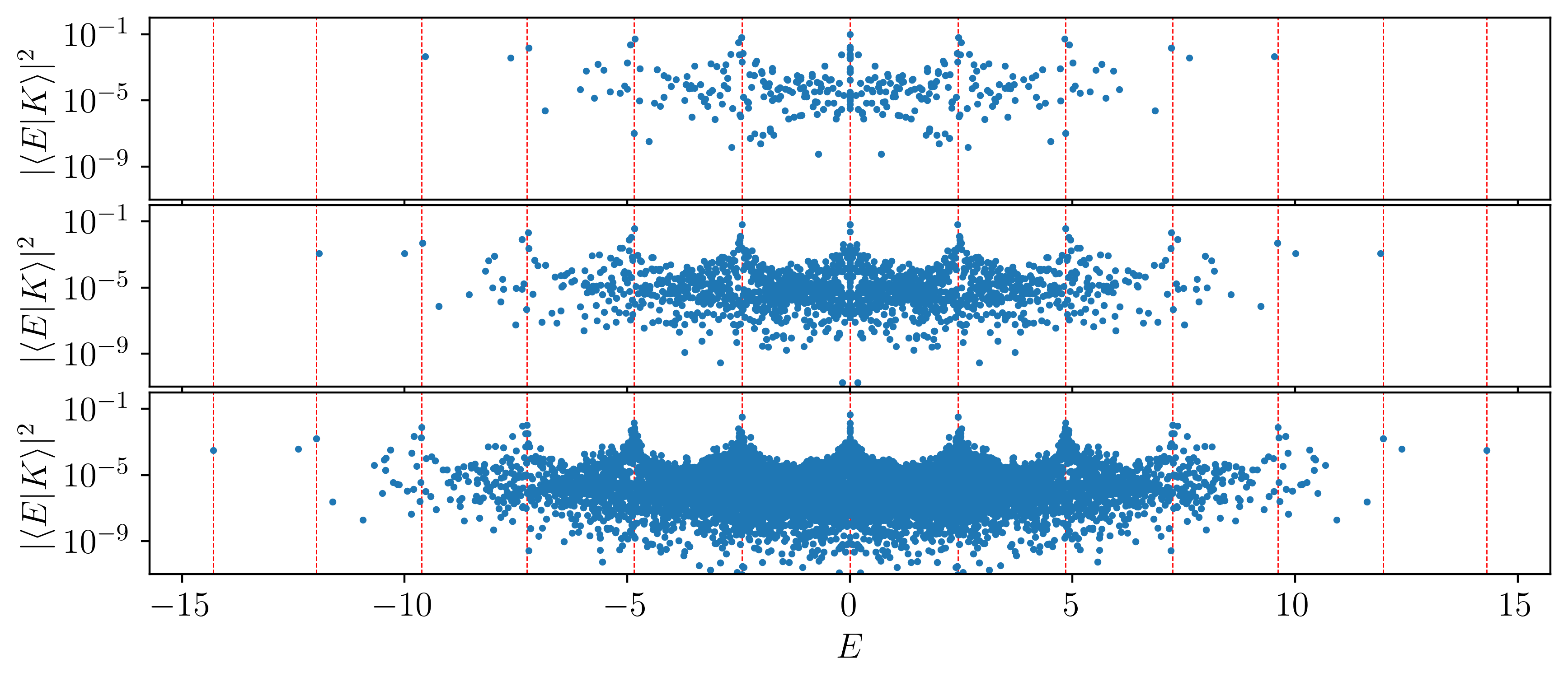}
\caption{Overlap of the $\ket{K=6}$ with the eigenstates of the chain with $\alpha=3$ and $N=24$, $30$ and $36$ (from top to bottom). The dashed vertical lines indicate the energy of the scarred states for $N=36$. The energy of the scarred states for the other two system sizes match well with that data.}
\label{fig:olap_scale}
\end{figure}

Beyond short-time revivals, one can also study the scaling of infinite-time quantities such as encoded in the eigenstates of the system. On Fig.~\ref{fig:olap_scale} we show the overlap of eigenstate with the $K=2\alpha$ initial state for $\alpha=3$ and three different system sizes. As mentioned in the main text, the number of towers of states grows as $1+2N/K$. Their energy spacing remains essentially constant with $N$, meaning that the period does not change as the system size is increased. While this is similar to what happens in the PXP model, the main difference with this model is the absence of a clear state in each tower with a significantly higher overlap than the others. Instead, the towers of state are much denser. This complicates the analysis of how the properties of a single scarred state evolve with $N$. Nonetheless, even in the PXP case, the scarred states start to hybridize with other eigenstates for large enough system sizes~\cite{TurnerPRB}. This leads to the same kind of phenomenology as we observe here for $\alpha>1$, with several eigenstates close to each other in overlap at the top of each tower. So for very large systems we expect the same type of behavior for all $\alpha$ investigated here as for the PXP model with $\alpha=1$.

\section{Quantum dimension}\label{SM:qdim}
For an arbitrary blockade radius $\alpha$, one can easily compute the number of states in the Hilbert space. The key is to separate all states according to the $\alpha$ leftmost states. States in the first class only have down-spins in these sites, states in the second class have one up-spin on the leftmost site, states in the third class have one up-spin on the second leftmost site and so on. As the blockade only allows at most one excitation on these $\alpha$ sites, the $\alpha+1$ classes encompass all possible states. Importantly, if we want to add a new site to the left of the chain the effect of the constraint on each of the classes is obvious. Let us consider a column-vector $\vec{v}_N$ of size $\alpha+1$ where entry $j$ is the number of states in class $j$ for $N$ sites (assuming OBC). Then if we glue a new site to the left while respecting the constraint, the new size vector $\vec{v}_{N+1}$ can be described using the transfer matrix $T_\alpha$ as 
\begin{equation}
\vec{v}_{N+1}=T_\alpha \vec{v}_{N}=T_\alpha^{N+1} \vec{v}_0 \quad \text{where} \quad
    T_\alpha=\begin{pmatrix}
        1 & 1 & 0 & 0&\ldots & 0 & 0 \\
        0 & 0 & 1 & 0 & \ldots & 0 &0 \\
        0 & 0 & 0 & 1 & \ldots & 0 &0 \\
        \vdots &  & & & \ddots   &  & \vdots \\
        0 & 0 & 0 & 0 & \ldots &  1 & 0 \\
        0 & 0 & 0 & 0 & \ldots &  0 & 1 \\
        1 & 0 & 0 & 0 & \ldots &  0 & 0 \\
    \end{pmatrix} \quad \text{and} \quad \vec{v}_0=\begin{pmatrix}
        1 \\
        0  \\
        0 \\
        \vdots \\
        0 \\
        0  \\
    \end{pmatrix}
\end{equation}
Here $T_\alpha$
is the transfer matrix of size $(\alpha+1) \times (\alpha +1)$. For PBC, a similar computation can be done. The number of states in class $j$ for size $N$ is simply given by $\left(T_\alpha^N\right)_{j,j}$, meaning that the size vector is the diagonal of $T_\alpha^N$. In both cases, the total number of states can be computed as the sum of the number of states in each class. A corollary of this is that the Hilbert space dimension grows as $d_\alpha^N$, where $d_\alpha$ is simply the largest magnitude eigenvalue of $T_\alpha$. 

Let us now compute this through the characteristic polynomial method that tells us that the eigenvalues of $T_\alpha$ are simply the roots of $\det(\lambda \mathbf{1}-T_\alpha)$. One can use the Laplace expansion with respect to the first column to find that
\begin{equation}
\det(\lambda \mathbf{1}-T_\alpha)=(\lambda-1)\det\begin{pmatrix}
        \lambda & -1 & 0 & 0&\ldots & 0 & 0 \\
        0 & \lambda & -1 & 0 & \ldots & 0 &0 \\
        0 & 0 & \lambda & -1 & \ldots & 0 &0 \\
        \vdots &  & & \ddots & \ddots   &  & \vdots \\
        0 & 0 & 0 & 0 & \ldots &  -1 & 0 \\
        0 & 0 & 0 & 0 & \ldots &  \lambda & -1 \\
        0 & 0 & 0 & 0 & \ldots &  0 & \lambda \\
    \end{pmatrix}-(-1)^\alpha \det\begin{pmatrix}
        -1 & 0 & 0&\ldots & 0 & 0 & 0 \\
        \lambda & -1 & 0 & \ldots & 0 & 0 &0 \\
        0 & \lambda & -1 & \ldots & 0 & 0 &0 \\
        \vdots &  & \ddots & \ddots &  &  & \vdots \\
        0 & 0 & 0 &  \ldots & -1& 0 & 0 \\
         0 & 0 & 0 & \ldots & \lambda & -1 & 0 \\
        0 & 0 & 0 & \ldots &  0& \lambda & -1 \\
    \end{pmatrix}.
\end{equation}
As both of the remaining matrices are triangular, their determinant can be computed easily. We thus end up with the characteristic polynomial $(\lambda-1)\lambda^\alpha-1$. As such, the $d_\alpha$ is equal to the largest magnitude solution of the equation $\lambda^{\alpha+1}=\lambda^\alpha+1$. We can immediately see that for the PXP case, this becomes $\lambda^2=\lambda+1$ which famously has for largest solution the golden ratio $\varphi=(1+\sqrt{5})/2$ as expected from previous works. We also note that $d$ for $\alpha=2$ is known as the supergolden ratio.

Alternatively, one can derive this scaling dimension for OBC by finding the recursion relation for the number of states $\mathcal{D}_N$ in a similar but more intuitive way. Let us consider gluing a site to the left of a chain with $N$ sites. We can always add a down-spin, giving us $\mathcal{D}_N$ configuration. However, we can only add an up-spin if the last $\alpha$ sites are down. But such configurations can always be obtained by gluing $\alpha$ down-spins to the left of a chain with $N-\alpha$ sites, meaning there are $\mathcal{D}_{N-\alpha}$ of them. We thus end up with the recursion relation $\mathcal{D}_{N+1}=\mathcal{D}_N+\mathcal{D}_{N-\alpha}$, which will give us the expected equation $\lambda^{\alpha+1}=\lambda^\alpha+1$ for the asymptotic scaling. Moreover, it shows how the system size can be computed using a number series similar to the Fibonacci sequence. 

\section{Additional fidelity and entanglement entropy data}\label{SM:data}
\begin{figure}
    \centering
\includegraphics[width=\columnwidth]{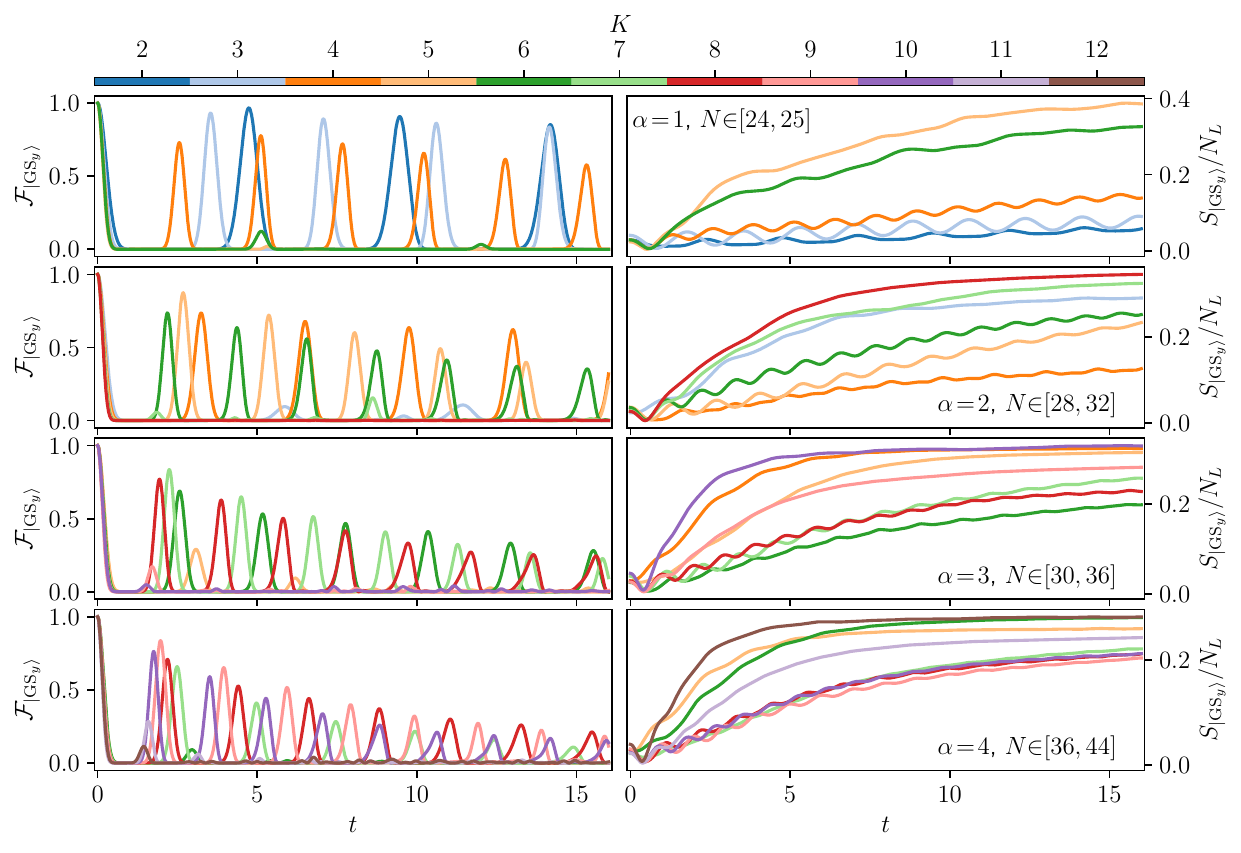}
\caption{Fidelity and entanglement entropy after a quench from $\GSy$. The entanglement entropy is divided by the size of the left subsystems in order to better compare different system sizes at later times.}
\label{fig:fid_quench}
\end{figure}

In this section, we provide additional data about the revivals in the various models. In Fig.~\ref{fig:fid_quench}, we show fidelity and entanglement entropy after a quench from the $\GSy$ state. From the fidelity, one can see that revivals with $K=2\alpha-1$ get progressively better as $\alpha$ is increased. For the entanglement entropy, a striking feature is that this quantity actually goes \emph{down} at short times
for $K=2\alpha$ to $2\alpha+2$.

\begin{figure}
    \centering
\includegraphics[width=\columnwidth]{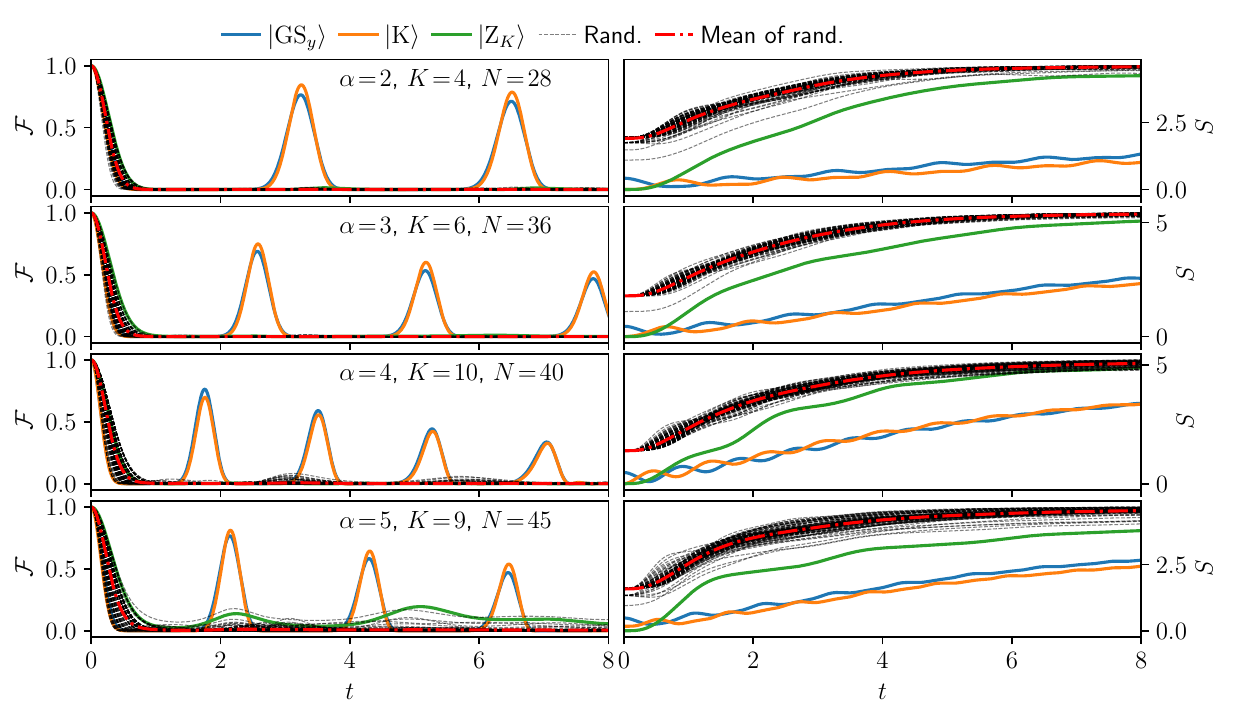}
\caption{Fidelity after a quench from various states. There are 100 random states in each panel. They are obtained by sampling basis states in the computational basis and projecting them onto the relevant momentum sectors. While the random states and the $\ket{\Zd{K}}$ states thermalize as expected, the $\ket{K}$ and $\GSy$ states show clear revivals and slow growth of entanglement entropy.}
\label{fig:fid_rand}
\end{figure}

In Fig.~\ref{fig:fid_rand}, we compare revivals from notable states to that of random states. Due to their $K$-site periodicity, the $\ket{K}$, $\GSy$ and $\Zd{K}$ states live in a reduced number of momentum sectors. In order to account for this, we select random initial states by uniformly drawing basis states and projecting them onto the same momentum sectors. 
For all blockade ranges investigated, we see a stark contrast between the behavior of the $\ket{K}$ and $\GSy$ states and that of other states. This is true both for fidelity and entanglement entropy. Nonetheless, this difference is less pronounced for $\alpha=4$ and $\alpha=5$. We attribute this to the need to go to much larger chains as the blockade range increases. Picking smaller systems for $\alpha=1,2,3$ leads to the same kind of behavior.

\section{Equations of motion in the semiclassical limit}\label{SM:eom}
In this section, we provide additional information regarding the semiclassical limit we obtain through our TDVP Ansatz in Appendix~\ref{app:TDVP}. We will focus on the cases $\alpha=1$, $K=4$ and $\alpha=2$, $K=4$. For each of them, we give the equations of motion (EOMs) for the $K$ angles $\theta_j$ as well as the expectation values of the occupation operator $n$ as a function of these angles. 

We only give these expression for the first site of the unit cell, as the corresponding expressions for the remaining sites are identical up to a cyclic shift of all indices $j\to (j\mod K)+1$.
\subsection{$\alpha=1$, $K=4$}
In this case the EOM for the first angle $\theta_1$ is
\begin{equation} \label{dotTheta}
\dot{\theta}_1{=}\cos(\theta_2)+ \cos(\theta_4)\sin(\theta_1)\sin(\theta_4)         \frac{\cos^2(\theta_3)+\cos^2(\theta_1)\sin^2(\theta_2)\sin^2(\theta_3)}{\cos^2(\theta_4)+ \cos^2(\theta_2)\sin^2(\theta_3)\sin^2(\theta_4)}.
\end{equation}

From the $\theta_j$ angles, we can directly compute the occupation of site 1 of the unit-cell as 
\begin{equation}
n_1=\frac{\sin(\theta_1)^2 \left[\cos(\theta_4)^2 + \cos(\theta_2)^2 \sin(\theta_3)^2 \sin(\theta_4)^2\right]}{1 - 
 \sin(\theta_1)^2 \sin(\theta_2)^2 \sin(\theta_3)^2 \sin(\theta_4)^2}.
\end{equation}

\subsection{$\alpha=2$, $K=4$}
The EOM for the first angle in the case with $\alpha=2$ and $K=4$ 
is
\begin{align}
\dot{\theta_1}&=\cos(\theta_3)\cos(\theta_2){+}\frac{\cos(\theta_4)\sin(\theta_1) }{\cos^2(\theta_3)\cos^2(\theta_4){+}\sin^2(\theta_3)\sin^2(\theta_2)}\times\\
&\left(\cos(\theta_3)\sin(\theta_3)\left[\sin^2(\theta_1)\sin^2(\theta_4){+}\cos^2(\theta_1)\cos^2(\theta_2)\right]{+}\cos(\theta_2)\sin(\theta_4)\left[\cos^2(\theta_2)\cos^2(\theta_3){+}\sin^2(\theta_1)\sin^2(\theta_2)\right]\right).\nonumber
\end{align}
The occupation on the first site of the unit cell is then
\begin{equation}
n_1=\frac{\sin(\theta_1)^2\left[\sin(\theta_2)^2 \sin(\theta_3)^2 + \cos(\theta_3)^2 \cos(\theta_4)^2\right]}{1 -\sin(\theta_2)^2 \cos(\theta_3)^2 \sin(\theta_4)^2 - 
 \sin(\theta_1)^2\cos(\theta_2)^2 \sin(\theta_3)^2  +\sin(
 \theta_1)^2 \sin(\theta_4)^2 \left[\sin(\theta_2)^2 + \sin(\theta_3)^2\right] }.
\end{equation}

\end{document}